\documentclass[pra,twocolumn,10pt,aps,superscriptaddress,longbibliography]{revtex4-2}
\setcitestyle{square}
\usepackage{graphicx}  
\usepackage{float}
\usepackage{dcolumn}   
\usepackage{bm}        
\usepackage{amssymb}   
\usepackage{amsmath}
\usepackage{isomath}
\usepackage[usenames, dvipsnames]{color}
\usepackage{soul}
\usepackage{physics}
\usepackage[normalem]{ulem}
\usepackage{xpatch}
\usepackage{IEEEtrantools}
\usepackage[newcommands]{ragged2e}
\usepackage{subcaption}
\usepackage{pgffor}
\usepackage{etoolbox}
\usepackage{xprintlen}
\usepackage{comment}
\usepackage{xcolor}
\usepackage{soul}
\usepackage[normalem]{ulem}

\usepackage{hyperref}
\hypersetup{
     colorlinks  = true,
     linkcolor    = red,
     citecolor    = cyan,
     urlcolor     = magenta
     }

\begin{document}
\widetext
\title{Remote Nucleation and Stationary Domain Walls via Transition Waves \\ in Tristable Magnetoelastic Lattices}
\author{Anusree Ray}
\thanks{These authors contributed equally}
\affiliation{Department of Aerospace Engineering, Indian Institute of Science, Bangalore, India, 560012}

\author{Samanvay Anand}
\thanks{These authors contributed equally}
\affiliation{Department of Aerospace Engineering, Indian Institute of Science, Bangalore, India, 560012}

\author{Vivekanand Dabade}
\email{Contact author: dabade@iisc.ac.in}
\affiliation{Department of Aerospace Engineering, Indian Institute of Science, Bangalore, India, 560012}

\author{Rajesh Chaunsali}
\email{Contact author: rchaunsali@iisc.ac.in}
\affiliation{Department of Aerospace Engineering, Indian Institute of Science, Bangalore, India, 560012}

\begin{abstract}
 We present a magnetoelastic lattice in which a localized external magnetic field, generated by an assembly of fixed magnets, tunes the potential landscape to create monostable, bistable, and tristable configurations. Focusing on the tristable potential, we numerically and experimentally confirm the existence of three distinct types of transition waves, each characterized by unique amplitudes and velocities, and establish a scaling law that governs their behavior. We also examine how these transition waves interact with the system's finite boundaries. Furthermore, by adjusting the potential symmetry through the localized external field, we investigate wave collision dynamics. In lattices with asymmetric potentials, the collision of similar transition waves leads to the remote nucleation of a third phase. In symmetric potentials, the collision of dissimilar transition waves results in the formation of a stationary domain wall, with its width tuned by the shape of the tristable potential well.
\end{abstract}
\maketitle
\section{Introduction}
Transition waves play a crucial role in various material processes, such as dislocation dynamics, where defects propagate through crystal lattices~\cite{braun2004frenkel}, and phase transitions in advanced materials such as ferroelectrics~\cite{Currie1979}, ferromagnets~\cite{Atkinson2003}, and shape memory alloys~\cite{falk1983ginzburg}. These waves are characterized by moving boundaries, referred to as phase boundaries or domain walls, which separate regions where the material exists in different phases. As the transition wave propagates, these boundaries shift, causing material elements to switch from one phase to another. Extensive theoretical studies have explored these waves in systems with nonconvex energy landscapes, where multiple stable equilibria exist, driving the sequential transition of elements from one equilibrium state to another~\cite{Remoissenet1984, Pouget1985, Abeyaratne1991, Peyrard1998, Flach1999, slepyan2004localized, Slepyan2005, truskinovsky2005kinetics, Benichou2013}.

Similar phenomena have emerged in metamaterials in recent years, where macroscopic mechanical systems exhibit multiple stable equilibria~\cite{nadkarni2016unidirectional, Deng2021}. By finely adjusting potentials and degrees of freedom, precise control over transition waves is achieved in metamaterials, enabling the design of reconfigurable structures with applications in soft robotics~\cite{Deng2020, Chi2022}, energy absorption~\cite{Shan2015}, deployable structures~\cite{Zareei2020}, and sound control~\cite{Ramakrishnan2020, Watkins2021}.

Initially, research into transition waves in metamaterials primarily focused on bi-stable unit cells with asymmetry sufficient to compensate for the system damping and make the transition wave sustainable~\cite{nadkarni2014dynamics, raney2016stable, katz2018solitary, shiroky2018propagation, Deng2020JMPS, zhouwang2023cooperative}. 
However, recent developments have expanded this design space by integrating magnetic elements into metamaterial lattices, enabling the realization of systems with more than two stable equilibria~\cite{yasuda2020transition, yasuda2023nucleation}. This has led to novel observations, such as the formation of stationary domain walls through the collision of two transition waves~\cite{yasuda2020transition}, a mechanism distinct from bistable lattices, which require additional defects to achieve similar outcomes~\cite{Kochmann2020, deng2020characterization}.
Furthermore, magnetic elements enable in situ tuning of the system’s potential energy landscape using an external magnetic field, without requiring physical changes to the lattice geometry. This provides more precise control over transition wave propagation and facilitates the creation of more adaptable and functional designs~\cite{pal2023programmable}.

Despite these advancements, a systematic design strategy for creating metamaterials with multiple stable equilibria using only external magnetic fields is still lacking. Such systems could allow for real-time tuning of the potential landscape—whether monostable, bistable, or tristable—without requiring changes to the metamaterial's geometry. This flexibility would support a wide range of wave dynamics. Notably, wave behavior in tristable configurations remains underexplored in the literature. Key open questions include: (1) How many types of transition waves can such lattices support, and is there a universal law governing their characteristics? (2) How do these waves interact with finite boundaries? (3) What happens when different transition waves collide?

In this work, we demonstrate how a localized external magnetic field, generated by an assembly of permanent magnets, can tune the potential landscape of a metamaterial lattice, enabling monostable, bistable, and tristable behavior. We focus on the tristable configuration, where experimental observations reveal the existence of three distinct transition waves. For the first time, we validate with experiments a scaling law relating wave velocity to power dissipation, as theoretically predicted in earlier studies~\cite{nadkarni2016universal}. Additionally, we investigate how boundary conditions affect these transition waves.

We further explore the collision dynamics of transition waves in both symmetric and asymmetric potentials, adjustable via the localized magnetic field. For asymmetric potentials, we show that two transition waves initiated from opposite ends collide, nucleating a third phase. Finally, for symmetric potentials, we experimentally demonstrate that the collision of two transition waves results in the formation of a stationary domain wall. Our findings reveal a previously unexplored correlation between the width of the domain wall and that of the colliding transition waves, governed by the shape of the tristable potential.

\begin{figure}
    \centering
    \includegraphics[width=\linewidth]{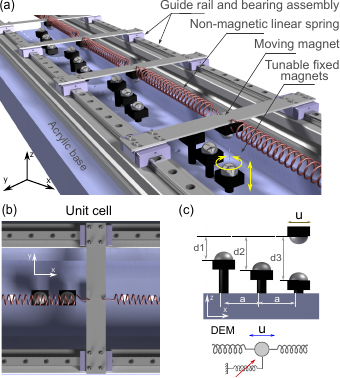}
    \caption{\justifying Experimental setup. 
(a) One-dimensional lattice assembly where the onsite potential is induced by fixed magnets. Moving masses are the sliding beams. 
(b) Top view of the unit cell, comprising three fixed magnets and one moving magnet attached beneath the sliding beam. 
(c) As per the discrete element model (DEM), the depths of the fixed magnets $(d_1, d_2,$ and $d_3)$ create a multi-stable onsite potential for the moving magnet.}
    \label{fig1}
\end{figure}

\section{Experimental setup}
We design a lattice comprising 10 unit cells, each consisting of a sliding beam (aluminum) connected to its neighbors via axial springs (phosphor-bronze), as depicted in Fig.~\ref{fig1}. The structure rests on a solid acrylic base supported by a pair of aluminum extrusions, housing the linear guide rail and bearing assembly (MGN7C, HIWIN). Underneath each beam is a Neodymium (N-52) permanent magnet, referred to as a ``moving magnet'', as it moves along the $x$-axis with the beam. Three magnets are affixed to the acrylic base for each unit cell, spaced at a distance $a=20$ mm between them. All the permanent magnets are spherical of a radius of $r=5$ mm and uniformly magnetized along the out-of-plane direction ($z$-axis). The unit cells are adequately spaced, with a distance $l= 120$ mm between them, ensuring that the moving magnets primarily interact with the three fixed magnets within their respective unit cells. By individually rotating the fixed magnets, we alter their depths ($d_1$, $d_2$, and $d_3$) and change the localized magnetic field, thus adjusting the effective onsite potential experienced by the moving magnet. For the measurements, we employ a Laser Doppler Vibrometer (Polytec single-point LDV) to detect the displacement of each moving mass.

\begin{figure}
    \centering
    \includegraphics[width=\linewidth]{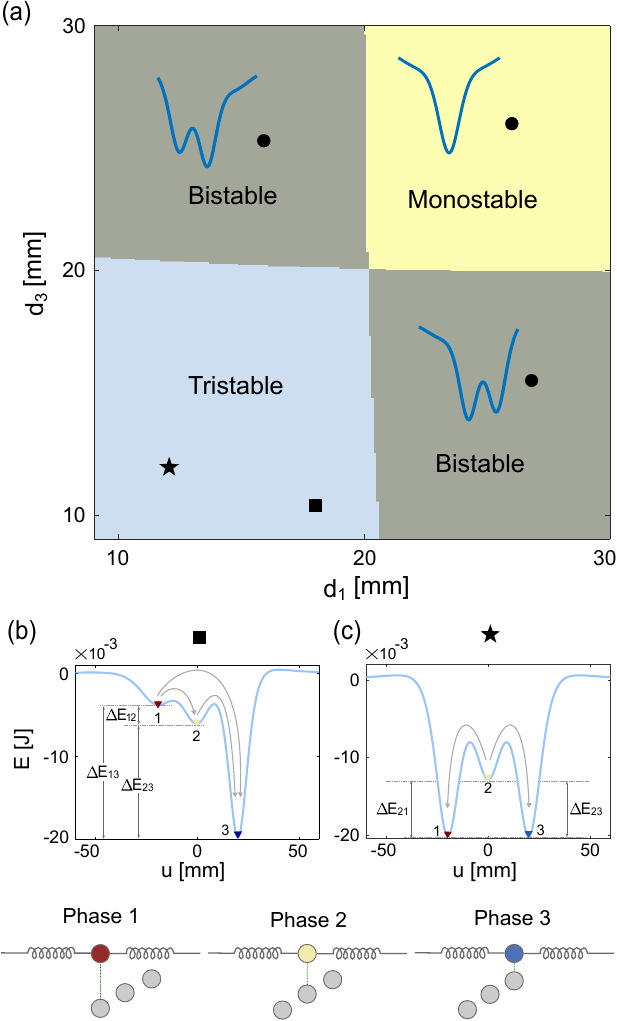}
    \caption{\justifying (a) Phase diagram of the onsite potential as a function of $d_1$ and $d_3$, for fixed $a= 20$ mm and ${d_2 = 15.3}$~mm. 
    (b) Asymmetric tristable onsite potential considered for the experiment, $(d_1, d_3) = (18, 10.4)$ mm. (c) Symmetric tristable onsite potential considered for the experiment $ (d_1, d_3) =  (12, 12)$ mm. Energy differences between different stable states are highlighted. Arrows denote different types of transition waves between \textit{Phase~1}, \textit{Phase 2}, and \textit{Phase 3}.}
    \label{fig2}
\end{figure}

\section{Numerical modeling}
\subsection{Nonlinear onsite potential}
{First, we model the magnetic interaction between moving and stationary magnets within a unit cell. Since all magnets possess uniform magnetization $m_s$ and are spherical with radius $r$, we derive the interaction energy, a function of $u$, the axial displacement of the moving mass, using Maxwell's equations (refer to Supplementary Material (Section I) for details):

\begin{equation}
\begin{split}
E(u) = & \frac{4 \pi m_s^2}{9 \mu_0} r^6 \Bigg[ \Bigg(\frac{1}{ ((u + a)^2 + d_1^2)^{\frac{3}{2}} }- \frac{3d_1^2}{((u + a)^2+ d_1^2)^{\frac{5}{2}}}\Bigg)\\
& + \left(\frac{1}{(u^2 + d_2^2)^{\frac{3}{2}}}- \frac{3d_2^2}{(u ^2+ d_2^2)^{\frac{5}{2}}}\right) \\
& + \left(\frac{1}{ ((u - a)^2 + d_3^2)^{\frac{3}{2}} }- \frac{3d_3^2}{((u - a)^2+ d_3^2)^{\frac{5}{2}}}\right)\Bigg],
\end{split}
\label{s1_eq1}
\end{equation}
where $\mu_0 = 4 \pi \times 10^{-7} \, \text{Vs/Am}$ denotes vacuum permeability. Magnetization of all N-52 magnets are considered as $m_s = 0.8 \,\text{T}$. Additionally, the depths of fixed magnets $d_1$, $d_2$, and $d_3$ are adjustable to modify the interaction energy landscape, as illustrated in Fig.~\ref{fig2}. We maintain $d_2$ constant and examine the energy landscape's behavior as $d_1$ and $d_3$ vary. Figure~\ref{fig2}(a) demonstrates the possibility of monostable, bistable, and tristable potential landscapes for various depth combinations. We focus on the tristable regime, presenting two distinct tristable landscapes, one asymmetric and the other symmetric about $u=0$, depicted in Figs.~\ref{fig2}(b) and \ref{fig2}(c) respectively. In the later sections, we will analyze transition waves that facilitate system switching across three different phases (corresponding to three local minima), namely \textit{Phase 1}, \textit{Phase 2}, and \textit{Phase 3}.

\subsection{Equations of motion}
We employ the discrete element method (DEM) to simulate the dynamics of our system. Sliding beams with moving magnets are treated as lumped masses interconnected by linear springs. To simplify calculations, we assume a significantly higher bending rigidity for the beams, neglecting out-of-plane motion along the $z$-axis for the moving mass.
The derivative of the nonconvex onsite energy, previously calculated, serves as a measure of the onsite force acting on the moving magnets. Consequently, the equation of motion for the $n$th unit cell is given by:

\begin{equation}
\begin{split}
& m u_{n,tt}  - k(u_{n-1} - 2u_n + u_{n+1}) + E'(u_n) \\
 & - c_1(u_{(n-1),t} - 2u_{n,t} + u_{(n+1),t}) 
 + c_2 \ \text{sgn}(u_{n,t}) = 0,
\end{split}
\label{s1_eq2}
\end{equation}
where $m$ represents the mass of the moving assembly, $k$ denotes the linear stiffness of the springs, $c_1$ and $c_2$ are intersite and onsite damping parameters, respectively. Variables following a comma in indices denote partial derivatives. As the moving mass slides on the guide rails, we utilize a dry friction damping model for the onsite term and a viscous damping model for the springs. 

\subsection{Scaling law for transition wave}
In this section, we derive the scaling law for transition waves propagating in the system with velocity $\nu$. We thus assume $u_n(t)=u(n l - \nu t) \equiv u(\xi)$ and substitute it into Eq.~\eqref{s1_eq2} to yield
\begin{align}
\begin{split}
m\nu^2u_{,\xi \xi}-k\big(u(\xi-l)-2u(\xi)+u(\xi+l) \big) \\
+ E'(u)
+ c_1\nu \big[u_{,\xi}(\xi-l) - 2u_{,\xi}(\xi)+u_{,\xi}(\xi+l) \big] \\
- c_2 \nu \ \text{sgn}({u_{,\xi}}) = 0.
\end{split}
\label{s3_eq1}
\end{align}
Equations are thus transformed into a traveling frame of reference $\xi$. Multiplying Eq.~\eqref{s3_eq1} by $u_{,\xi}$ and integrating over the real $\xi$ axis, we obtain
\begin{align}
\begin{split}
\int_{-\infty}^{\infty} \bigg[m\nu^2u_{,\xi \xi}-k\big(u(\xi-l)-2u(\xi)+u(\xi+l) \big) \\
+ E'(u) \
+c_1 \nu \big(u_{,\xi}(\xi-l) - 2u_{,\xi}(\xi)+u_{,\xi}(\xi+l) \big) \\
- c_2 \nu \ \text{sgn}({u_{,\xi}}) \bigg]u_{,\xi}d\xi=0.
\end{split}
\label{s3_eq2}
\end{align}
If the transition wave changes the system from the initial phase $u_i$ at ${t \rightarrow -\infty}$ (${\xi \rightarrow \infty}$) to $u_f$ at ${t \rightarrow \infty}$ (${\xi \rightarrow -\infty}$), and hence we impose ${u(\xi \rightarrow \infty) = u_i}$ and ${u(\xi \rightarrow -\infty) = u_f}$. Furthermore, for a dissipative system, the wave profile would reach a steady state at ${t \rightarrow \infty}$, implying ${u_{,\xi} (\xi \rightarrow -\infty) = 0}$. Since the system was initially at rest, we also have ${u_{,\xi} (\xi \rightarrow \infty) = 0}$. 
Upon examining the individual integrals in Eq.~\eqref{s3_eq2}, we find
\begin{equation}
\int_{-\infty}^{\infty} \left(m\nu^2u_{,\xi \xi}\right)u_{,\xi}d\xi= \int_{-\infty}^{\infty} \dfrac{m\nu^2}{2} \dfrac{d}{d\xi}\big(u_{,\xi}\big)^2 d\xi=0.
\label{s3_eq3}
\end{equation}
Next we compute the integral 
$$\mathcal{I} =  \int_{-\infty}^{\infty} \left[k\left(u(\xi-l)-2u(\xi)+u(\xi+l) \right)\right]u_{,\xi}d\xi. $$
The second term of $\mathcal{I}$ reduced as
\begin{align}
-2k\int_{-\infty}^{\infty}  u(\xi)u_{,\xi}d\xi \nonumber \\
&\hspace{-1cm} = -k\int_{-\infty}^{\infty} \dfrac{d}{d\xi}  \big(u({\xi})\big)^2 d\xi = -k\big(u_i^2-u_f^2\big). 
\label{s3_eq4a}
\end{align}
We define $\eta = \xi-l$ and subsequently the first term of $\mathcal{I}$ can be rewritten as
\begin{equation}
\int_{-\infty}^{\infty} ku(\xi-l)u_{,\xi} (\xi) d\xi =  \int_{-\infty}^{\infty} ku(\eta)u_{,\eta}(\eta+l)d\eta.
\label{s3_eq4b}
\end{equation}
Since, $\eta$ is a dummy variable, this can be rewritten as
\begin{equation}
\int_{-\infty}^{\infty} ku(\eta)u_{,\eta}(\eta+l)d\eta= \int_{-\infty}^{\infty} ku(\xi)u_{,\xi}(\xi+l)d\xi .
\label{s3_eq4c}
\end{equation}
Therefore, the first and the third terms of $\mathcal{I}$ are deduced to 
\begin{align}
\int_{-\infty}^{\infty}  & \left[ k\left(u(\xi-l)  + (\xi+l) \right) \right] u_{,\xi}d\xi  \nonumber \\  
& = k \int_{-\infty}^{\infty} \left[ u(\xi) u_{,\xi} (\xi+l) +u(\xi+l) u_{,\xi}(\xi) \right] d\xi  \nonumber \\
& = k \int_{-\infty}^{\infty} \dfrac{d}{d \xi} \left[ u(\xi) u(\xi+l) \right] = k \big(u_i^2-u_f^2 \big).
\label{s3_eq4d}
\end{align}
Combining Eqs.~\eqref{s3_eq4a} and ~\eqref{s3_eq4d}, the integral $\mathcal{I} = 0$ .
Finally, the term in Eq.~\eqref{s3_eq2}
\begin{equation}
    \int_{-\infty}^{\infty} E'(u) u_{,\xi}d\xi= E(u_i)-E(u_f)= \Delta E.
    \label{s3_eq5}
\end{equation}
Therefore Eq.~\eqref{s3_eq2} reduces to 
\begin{align}
\begin{split}
\Delta E =  \int_{-\infty}^{\infty} \bigg[   &- c_1 \nu \big(u_{,\xi}(\xi -l ) - 2u_{,\xi}(\xi)+u_{,\xi}(\xi+l) \big) \\
& + c_2 \nu \ \text{sgn}({u_{,\xi}}) \bigg]u_{,\xi}d\xi.
\end{split}
\label{s3_eq5a}
\end{align}
Multiplying both sides with $\nu/l$, approximating the right-hand side by a discrete system, we obtain
\begin{align}
\begin{split}
{\nu \Delta e} \simeq  \sum_{n=1}^{N} \bigg[   & - c_1  \big(u_{n-1, t} - 2u_{n,t}+u_{n+1, t} \big) \\
& + c_2 \ \text{sgn}({u_{n,t}}) \bigg]  u_{n,t	} \equiv P_d.
\end{split}
\label{s3_eq6}
\end{align}
where  $\Delta e \equiv \Delta E / l$ represents the change in onsite energy per unit length. The right-hand side denotes the total power dissipated ($P_d$) by viscous and Coulomb damping. Due to discreteness, $P_d$ oscillates in time with a period $T = l/\nu$ (see Supplementary Material (Section II) for details). Therefore, we compute the time-averaged $P_d$ as
\begin{equation}
\langle P_d \rangle
= \frac{1}{T} \int_{t_0}^{t_0 + T}  P_d \ dt,
\label{s3_eq7}
\end{equation}
and we modify the scaling law as
\begin{equation}
\nu  \Delta e  \simeq \langle P_d \rangle.
\label{s3_eq8}
\end{equation}

A similar scaling law can be found in the work of Neel et al.~\cite{nadkarni2016universal}. However, we have validated this prediction for the simultaneous presence of nonlinear onsite damping and linear intersite damping in our system. Since the  power dissipated on the right-hand side is always positive due to the second law, we can conclude that $\nu \Delta e \geq 0$, which is analogous to the entropy condition derived in Ref.~\cite{Abeyaratne1991}, with $\Delta e$ acting as the driving force on the phase boundary propagation.
In our tristable lattice, the scaling remains independent of inter-particle stiffness. Furthermore, the specific topology of the onsite potential does not affect the scaling law; instead, it depends on the initial and final configuration of the energy state. 

\begin{figure*}[!]
    \centering
    \includegraphics[width=\linewidth]{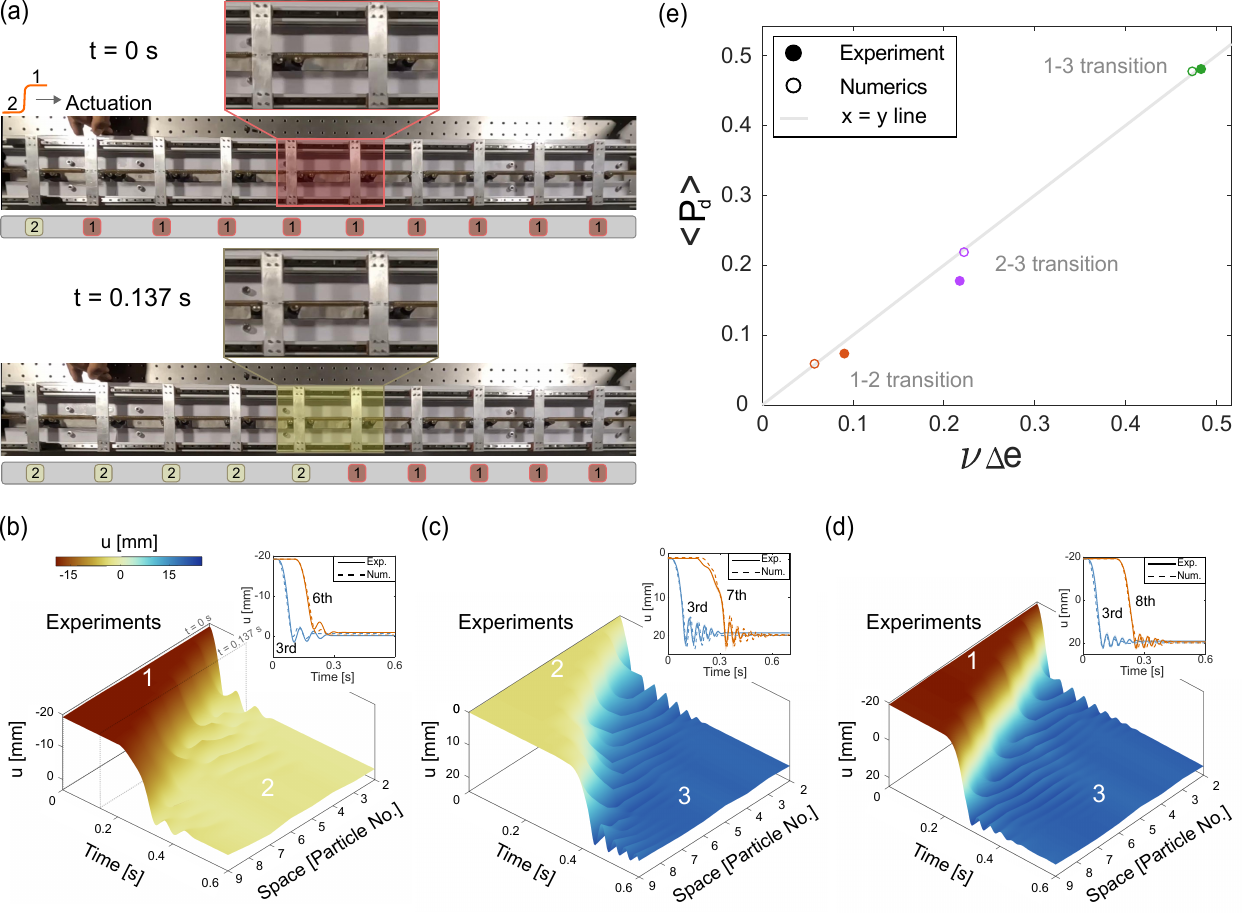}
    \caption{\justifying Three types of transition waves.
     (a) Snapshots from experiments demonstrating the $1 \rightarrow 2$ transition wave. The unit cells are labeled with the phases. 
     (b),(c),(d) Spatiotemporal plots of displacement obtained from experiments for $1 \rightarrow 2$, $2 \rightarrow 3$, and $1 \rightarrow 3$ transition waves, respectively. 
     Insets show the displacement time series of different moving masses. 
     (e) Experimental validation of the scaling law in Eq.~\eqref{s3_eq8}.
     }
    \label{fig3}
\end{figure*}

\section{Results and Discussions}

In this section, we focus on tristable configurations and present both numerical and experimental results addressing four key aspects: (A) the formation of transition waves, (B) the influence of boundary conditions, (C) nucleation behavior, and (D) the existence of stationary domain walls.
 
\subsection{Formation of transition waves}
We perform experiments on a chain comprising 10 unit cells with an asymmetric tristable onsite potential, as depicted in Fig.~\ref{fig2}(b). 
Due to asymmetry in the well, we expect several sustainable transition waves propagating in the lattice. These wave profiles are also referred to as \textit{kinks} or \textit{antikinks}~\cite{remoissenet2013waves}, and manifest as localized wave packets propagating with a constant shape. We identify three distinct types of transition waves: (i) $1 \rightarrow 2$, with the transition from \textit{Phase 1} to \textit{Phase 2}; (ii) $2 \rightarrow 3$, with the transition from \textit{Phase 2} to \textit{Phase 3}; and (iii) $1 \rightarrow 3$, with the transition from \textit{Phase 1} to \textit{Phase 3}. 
Note that transition waves, such as $3\rightarrow 1$, $3\rightarrow 2$, or $2\rightarrow 1$, are energetically not favorable as the final state is at higher energy, which enhances the effect of dissipation and leads to the absence of sustainable transition wave~\cite{nadkarni2016unidirectional}.

First, we consider a scenario where all unit cells are in the highest energy state, i.e., \textit{Phase 1}, and both ends of the chain are fixed. 
We rely on pre-straining the first spring and releasing its strain energy so that we can initiate the transition wave in a controlled and repeatable manner. The pre-strain is introduced by snapping the first unit cell into \textit{Phase 2} and fixing it in place while restraining the second unit cell, causing the spring between them to develop compressive strain. When the second unit cell is released, the transition wave is initiated.
We observe a large-amplitude nonlinear wave propagating through the lattice, transitioning each unit cell from \textit{Phase 1} to \textit{Phase 2} as shown in Fig.~\ref{fig3}(a). The spatiotemporal map of the displacement is plotted in Fig.~\ref{fig3}(b), clearly indicating the propagation of the $1 \rightarrow 2$ transition wave.
Similarly, we conduct experiments and observe $2 \rightarrow 3$ and $1 \rightarrow 3$ transition waves as shown in Fig.~\ref{fig3}(c) and \ref{fig3}(d), respectively. 
We also conduct numerical simulations on longer chains with 100 unit cells (see Supplementary Material (Section II) for details). We confirm that all types of transition waves are sustainable in these longer chains. 
This further demonstrates the fact that the asymmetries in the potential wells lead to energy gain after each snapping of unit cells, which compensates for the energy lost due to damping and thus facilitates a sustainable propagation of transition waves ~\cite{raney2016stable, yasuda2020transition}. 

Furthermore, experimental data is utilized to fine-tune the damping parameters of the numerical model, with values of $c_1$ as 0.05, 0.1, and 0.001 N s/m, and $c_2$ as 0.11, 0.13, and 0.2 N for $1 \rightarrow 2$, $2 \rightarrow 3$, and $1 \rightarrow 3$ transition waves, respectively. The inset of Fig.~\ref{fig3}(b)--\ref{fig3}(d) shows the comparison of temporal dynamics measured experimentally and modeled numerically.

From Fig.~\ref{fig3}(b)--\ref{fig3}(d), we observe that the $1 \rightarrow 2$, $2 \rightarrow 3$, and $1 \rightarrow 3$ transition waves reach the end of the chain at approximately 0.32 s, 0.42 s, and 0.27 s, respectively. This means that the $1 \rightarrow 3$ transition wave has the highest velocity, while the $2 \rightarrow 3$ transition wave has the lowest. This can be explained using the scaling law derived in Eq.\eqref{s3_eq8}, which relates the velocity of a transition wave ($\nu$) to the dissipated power ($\langle P_d \rangle$) and the energy difference ($\Delta e$) in the onsite potential. In Fig.~\ref{fig3}(e), we plot $\langle P_d \rangle$ against $\nu \Delta e$ from both experimental and numerical results. The data points lie on a straight line with slope 1, confirming the validity of the scaling law.
In our system, the power dissipated by the different transition waves follows the order: $\langle P_d \rangle_{1 \rightarrow 3} > \langle P_d \rangle_{2 \rightarrow 3} > \langle P_d \rangle_{1 \rightarrow 2}$. However, the energy difference also follows the relation: $ \Delta e_{1 \rightarrow 3} > \Delta e_{2 \rightarrow 3} > \Delta e_{1 \rightarrow 2}$. Consequently, the wave velocity ($\nu$), which depends on the ratio $\langle P_d \rangle / \Delta e$, follows the order $ \nu_{1 \rightarrow 3} > \nu_{1 \rightarrow 2} > \nu_{2 \rightarrow 3}$ for this particular setup.

\begin{figure}[t!]
    \centering
    \includegraphics[width=\linewidth]{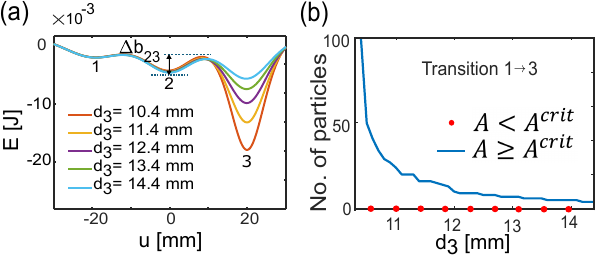}
    \caption{\justifying 
    (a) Onsite potential for varying values of $d_3$ keeping $d_1= 18$ mm and $d_2=15.3$ mm, fixed. 
    (b) Number of particles undergoing the $1 \rightarrow 3$ transition in a chain of length 100 vs $d_3$. The $1 \rightarrow 3$ transition is dependent on the initial amplitude $A$. For all initial amplitudes less than $A^{crit}$ (= 32 mm), we do not observe the $1 \rightarrow 3$ transition wave.  
    } 
    \label{fig4}
\end{figure}

To the best of the authors' knowledge, the $1 \rightarrow 3$ transition wave, which progresses directly from \textit{Phase 1} to \textit{Phase 3} without stabilizing at the intermediate state, has not been previously documented in the literature. 
We further investigate this unique transition wave by varying the onsite potential and excitation amplitude.
First, we keep $d_1$ and $d_2$ constant while varying $d_3$ to control the energy of \textit{Phase 3} relative to \textit{Phases 1} and \textit{2}, while maintaining nearly constant energy barriers, as shown in Fig.~\ref{fig4}(a). As $d_3$ increases, the third well becomes shallower, reducing the asymmetry and lowering the energy differences $\Delta E_{13}$ and $\Delta E_{23}$.
Next, we conduct numerical simulations on a system of 100 unit cells with varying $d_3$ to examine the propagation length of the transition wave. When the initial amplitude exceeds the critical threshold to cross the energy barrier between \textit{Phase 2} and \textit{Phase 3} ($A^{\text{crit}} = 32$ mm), we observe the $1 \rightarrow 3$ transition wave propagating through the system as shown in Fig.~\ref{fig4}(b). However, for larger values of $d_3$, fewer particles participate in the wave propagation, suggesting that reduced asymmetry of \textit{Phase 3} with \textit{Phases 1} and \textit{2} results in lower energy gain with each snap, thereby shortening the propagation distance.

\subsection{Effect of boundary conditions}
In this section, we evaluate the role of boundary conditions on the behavior of transition waves. All of our previous studies have been conducted on chains with fixed-fixed boundary conditions, where we observed no reflection of transition waves upon reaching the opposite end of the chain.

To assess whether this observed behavior persists under varying boundary conditions, we analyze a chain of 100 unit cells with a distinct stiffness, $k_b$, applied only to the final spring in the chain, as depicted in Fig.~\ref{fig5}(a). The inter-particle stiffness for all other springs remains $k$. We conduct a numerical study on the reflection of transition waves as a function of $k_b$. For instance, $k_b=0$ corresponds to a free boundary condition. The spatiotemporal plot of displacement for this case is shown in Fig.~\ref{fig5}(b). We observe a $1 \rightarrow 2$ transition wave propagating from left to right, converting the entire chain into \textit{Phase 2}. Upon reaching the free boundary at the right end, a reflected $2 \rightarrow 3$ transition wave initiates, converting the chain from \textit{Phase 2} to \textit{Phase 3}, the lowest energy state. Notably, no reflection occurs when $2 \rightarrow 3$ or $1 \rightarrow 3$ transition waves approach the free boundary, as the chain has already reached its most stable state (\textit{Phase 3}) by this point, with no lower energy states available.

\begin{figure}[t!]
    \centering
    \includegraphics[width=\linewidth]{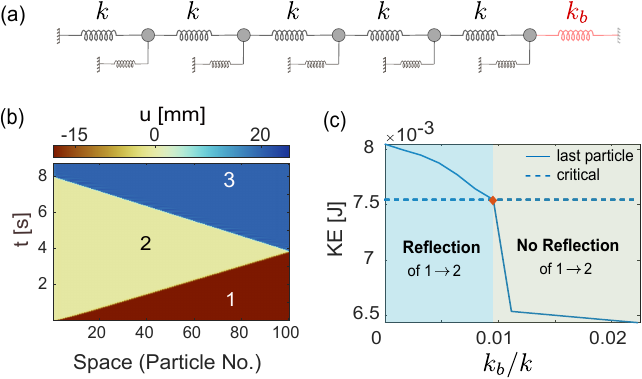}
    \caption{\justifying 
    (a) Schematic of a chain with different stiffness $(k_b)$ at the right boundary.
    (b) Spatiotemporal plot of displacement indicating reflection of $1\rightarrow 2$ transition wave from the free (right) end. (c) Kinetic energy (KE) vs normalized stiffness. The red marker represents the maximum normalized stiffness $(k_b/k)$ at which a reflected transition wave can occur. The blue and green regions indicate the range of stiffness corresponding to the presence or absence of the reflected transition wave.}
    \label{fig5}
\end{figure}

Next, we investigate wave reflection in chains with varying values of $k_b$. Our results show that the ${1 \rightarrow 2}$ transition wave reflects as a $2 \rightarrow 3$ wave when ${k_b/k < 0.0096}$. This threshold is explained by tracking the peak kinetic energy of the last particle, comparing it to the effective energy barrier necessary to transition to the lower energy state, \textit{Phase 3}. Figure~\ref{fig5}(c) illustrates the peak kinetic energy as a function of the normalized stiffness $k_b/k$ alongside the critical kinetic energy required for the $2 \rightarrow 3$ transition. This critical kinetic energy, derived from simulations of the $2 \rightarrow 3$ wave in the chain’s bulk under identical parameters, is given as $KE_{{2\rightarrow 3}}^{^{crit}}= 3.18 \times \Delta b_{23}$, where $\Delta b_{23}$ represents the energy barrier between \textit{Phase 2} and \textit{Phase 3}, with the factor of 3.18 capturing inter-particle stiffness and onsite potential asymmetry effects. Our findings indicate that reflection occurs when the peak kinetic energy of the last particle surpasses this critical threshold. These insights open new avenues for manipulating the final state of a multistable lattice through boundary condition adjustments.

\begin{figure*}[!]
\centering
    \includegraphics[width=\linewidth]{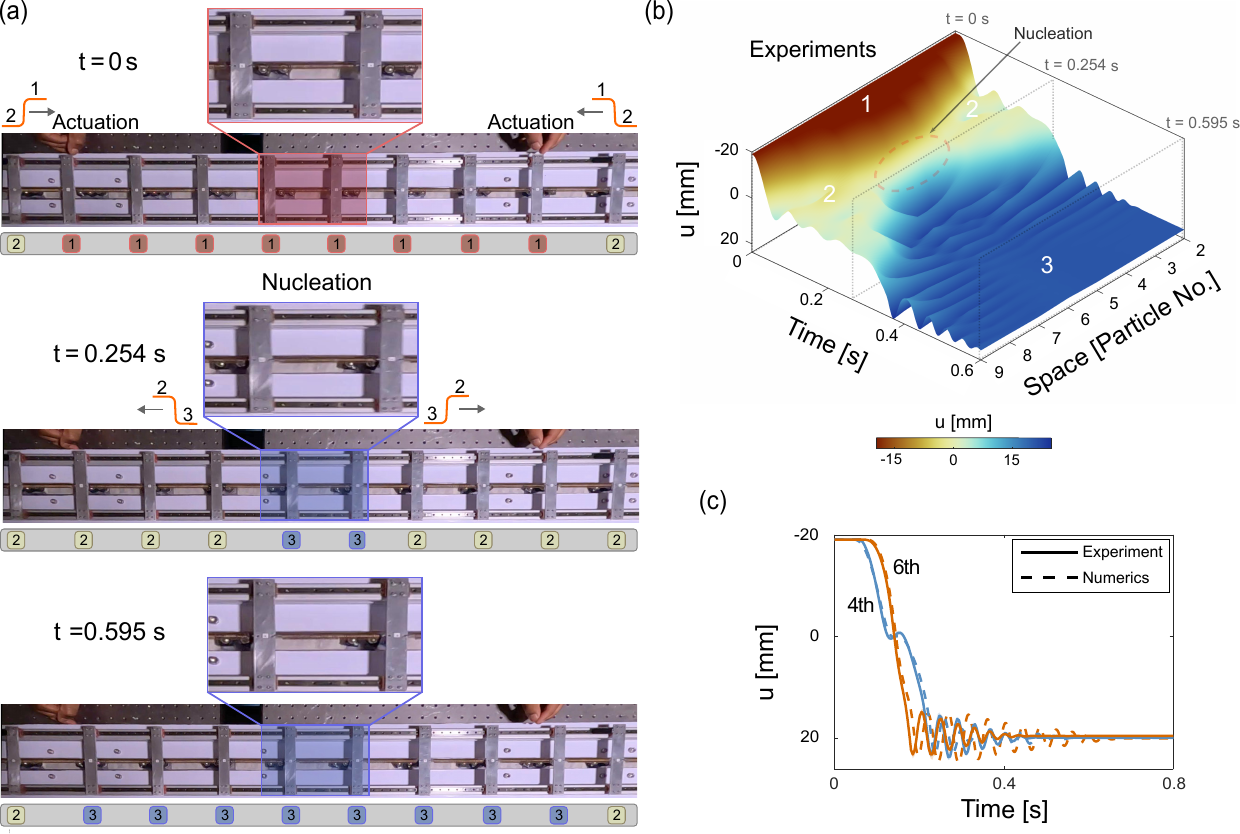}
    \caption{\justifying Collision of two $1 \rightarrow 2$ transition waves nucleates a new phase, \textit{Phase 3}. 
    (a) Experimental snapshots highlighting the excitation and nucleation. The unit cells are labeled with the phases. 
     (b) Spatiotemporal plot obtained from experiments showing the nucleation phenomena. The nucleus is formed at the 5th and 6th particles. 
     (c) Displacement time series for the 4th and the 6th particles.} 
    \label{fig6}
\end{figure*}

\subsection{Nucleation}
In this section, we study the collision of two transition waves. We consider a chain with an asymmetric tristable onsite potential shown in Fig.~\ref{fig2}(b) and excite the chain from both ends. Initially, the entire chain is in \textit{Phase~1}. We then trigger $1 \rightarrow 2$ transition waves from both ends. We observe transition waves propagating towards each other from the extreme ends and colliding at the middle of the chain at about $t=0.254$ s, as shown in Fig.~\ref{fig6}(a). The collision induces larger displacements and thereby nucleates a new phase, i.e., \textit{Phase 3}, in the 5th and 6th unit cells. Consequently, this nucleus triggers $2 \rightarrow 3$ transition waves from the middle of the chain that propagate back to the boundaries.

In Fig.~\ref{fig6}(b), we show an experimentally measured spatiotemporal map of displacement. The formation of a remote nucleus and the lattice transforming to \textit{Phase 3} is evident. We further show the temporal dynamics of the 4th and 6th unit cells in Fig.~\ref{fig6}(c). We observe that the 4th unit cell transitions to \textit{Phase 2} ($u \approx 0$) before the 6th unit cell. However, the latter transitions to \textit{Phase 3} earlier than the 4th unit cell. This is consistent with the earlier observation that nucleation occurs at the 5th and 6th unit cells.

\begin{figure}
    \centering
    \includegraphics[width=\linewidth]{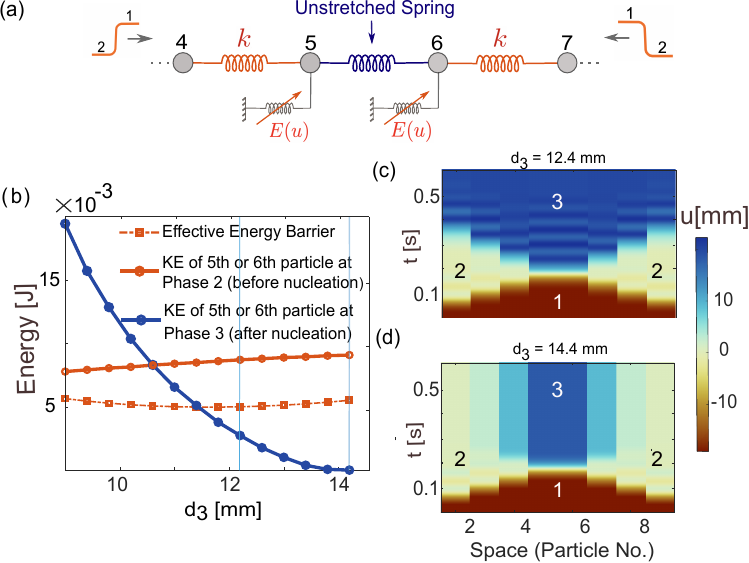}
    \caption{\justifying 
    (a) Schematic of the chain highlighting the region of collision of two $1 \rightarrow 2$ transition waves. 
    (b) Energy vs $d_3$. The effective energy barrier is compared with the instantaneous kinetic energy of the 5th or 6th particles when they reach \textit{Phase 2}. Moreover, the instantaneous kinetic energy of 5th or 6th particles when they reach \textit{Phase 3} dictates the propagation after nucleation (blue curve). The vertical lines represent two cases: Case (i) $d_3= 12.4$ mm and Case (ii) $d_3= 14.4$ mm, respectively. 
    (c) and (d) Spatiotemporal plot of displacement for cases (i) and (ii), respectively.
    } 
    \label{fig7}
\end{figure}

\begin{figure*}[!]
\centering
    \includegraphics[width=\linewidth]{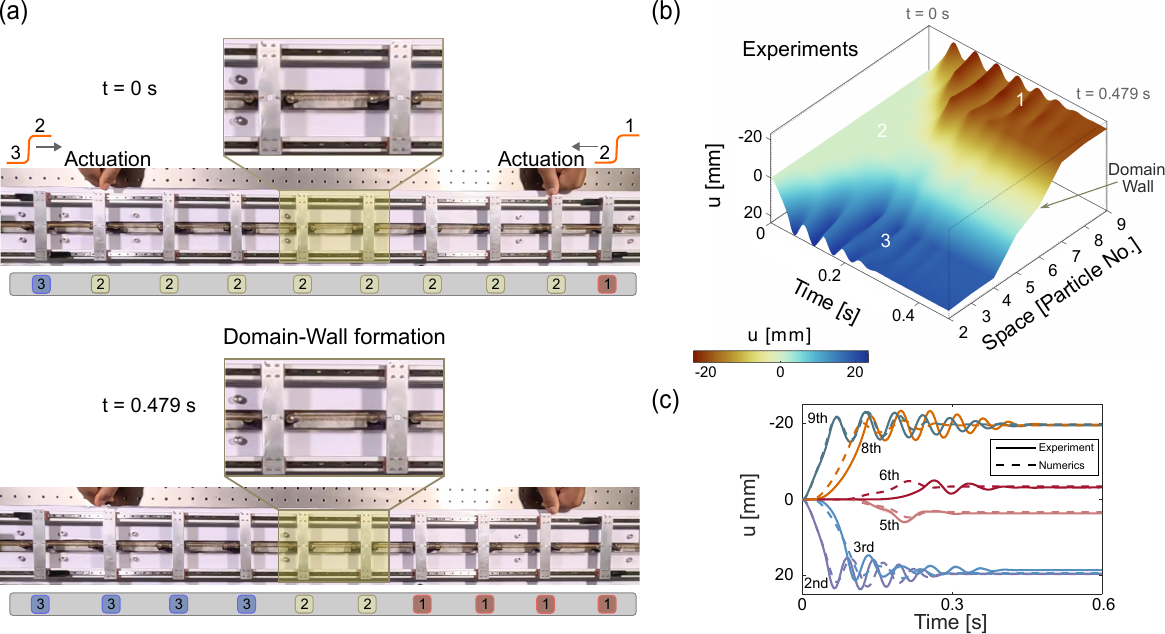}
    \caption{\justifying Collision of two transition waves leads to a stationary domain wall. 
    (a) Experimental snapshots highlighting the formation of the stationary domain. The unit cells are labeled with the phases.
    (b) Spatiotemporal plot obtained from experiments showing the formation of the stationary domain wall.
    (c) Displacement time series of several particles in the chain.}
    \label{fig8}
\end{figure*}

Nucleation can be interpreted as the collision of a kink and an anti-kink traveling in opposite directions~\cite{remoissenet2013waves}. In our study, the asymmetric tristable potential ensures that the propagation of these kinks and anti-kinks remains sustainable, even in longer chains and under damping (see Supplementary Material (Section III)). 
The experimental demonstration of this nucleation mechanism is a novel aspect of our work. Previous studies explored nucleation via collisions of vector solitons or breathers~\cite{jiao2024phase, PALIOVAIOS2024102199, yasuda2023nucleation}, which relied on only two stable states. In contrast, our mechanism leverages the design of all three stable states. Furthermore, the sustainability of transition waves in larger systems under damping highlights the robustness of this approach for achieving nucleation in extensive systems.

Next, we delve deeper into the nucleation mechanism to assess its resilience under the varying asymmetry of the tristable onsite potential. By keeping \(d_1\) and \(d_2\) constant and varying \(d_3\), as shown in Fig.~\ref{fig4}(a), we simulate the collision of ${1 \rightarrow 2}$ transition waves in a 10-particle chain. The focus is on the central particles (5th and 6th), which must acquire sufficient kinetic energy to overcome the effective energy barrier between \textit{Phase~2} and \textit{Phase~3}.
During the collision, the spring connecting the 5th and 6th particles remains unstretched, allowing the effective energy barrier to be approximated using the effective potential. This potential combines the onsite energy \((E(u))\) of the 5th (or 6th) particle with the intersite energy \((\frac{1}{2} k u^2)\) of the spring connecting it to the 4th (or 7th) particle, as shown in Fig.~\ref{fig7}(a) and given as
\begin{equation}
    \text{Energy}_\text{effective}^{2\rightarrow 3} = E(u) + \frac{1}{2} k u^2.
\end{equation}
The effective energy barrier is determined by calculating the local maximum of the effective potential. This maximum occurs at \( u_{min}^{2\rightarrow 3} \approx 10.712~\text{mm} \), a point between \textit{Phase~2} and \textit{Phase~3}. The barrier is then given as
\begin{equation}
    \text{Barrier}_\text{effective}^{2\rightarrow 3} =  \left| E(u_{min}^{2\rightarrow 3}) -  E(\text{Phase 2}) \right| + \frac{1}{2} k \left( u_{min}^{2\rightarrow 3} \right)^2.
\end{equation}

Figure~\ref{fig7}(b) shows the variation of the effective energy barrier with \(d_3\). Additionally, the kinetic energy of the 5th and 6th particles is plotted at the moment they reach \textit{Phase~2}. Remarkably, the kinetic energy consistently exceeds the effective energy barrier, confirming that the 5th and 6th particles successfully reach \textit{Phase~3}, thereby initiating nucleation for all tested values of $d_3$.

We also track the kinetic energy of the 5th or 6th particles immediately after nucleation, i.e., when they reach \textit{Phase~3}, as shown in Fig.~\ref{fig7}(b) as a function of \(d_3\). For lower values of \(d_3\) (corresponding to a deeper third well), this post-nucleation kinetic energy is significantly high, enabling the nucleation to propagate as a \(2 \rightarrow 3\) transition wave toward the boundaries. 
However, for higher values of \(d_3\), where \textit{Phase 2} and \textit{Phase 3} approach similar energy levels, nucleation occurs but fails to propagate toward the boundaries of the chain. This failure arises from the reduced kinetic energy of the nucleating particles post-nucleation for large \(d_3\).

To illustrate the effect of \(d_3\) on the occurrence and propagation of nucleation, we present spatiotemporal displacement maps for two cases: \(d_3 = 12.4\)~mm and \(d_3 = 14.4\)~mm, in Fig.~\ref{fig7}(c) and Fig.~\ref{fig7}(d), respectively. The former demonstrates both nucleation and propagation, while the latter (with larger \(d_3\)) shows nucleation but no subsequent propagation. Nonetheless, nucleation and propagation are observed over a wide range of asymmetries between the stable states, as dictated by \(d_3\).

Interestingly, the size of the nucleus depends on the number of particles in the chain. For example, two particles (the 5th and 6th) create the nucleus in the case above. Moreover, for an odd number of particles in the chain, it is possible to have a nucleus of only one particle (see Supplementary Material (Section III)). This implies that even if only a single particle is nucleated due to the collision of a kink and an anti-kink, it can effectively induce the propagation of the new phase in both directions. Moreover, it is also possible to remotely nucleate a new phase at an arbitrary location (and not only at the center) in the chain using a time delay of actuation from either end~\cite{jiao2024phase}. Refer to Supplementary Material (Section III) for details.

\subsection{Stationary Domain Wall}
In this section, we investigate the collision of two transition waves but with different onsite potentials. The tunability of our magnetoelastic lattice allows us to obtain a symmetric onsite potential well, as shown in Fig.~\ref{fig2}(c). The difference in the energy levels of \textit{Phase~2} with \textit{Phase~1} and \textit{Phase~3} enables two different types of transition waves in the system. Initially, the whole lattice is kept in \textit{Phase~2}. We trigger $2 \rightarrow 1$ and $2 \rightarrow 3$ transition waves from opposite ends and observe their collision. This scenario can also be understood as a collision of two kinks~\cite{remoissenet2013waves}.

In Fig.~\ref{fig8}(a), we observe two distinct propagating transition waves (moving domain walls) that collide at the center of the chain. However, the 5th and the 6th particles in the chain continue to remain nearly in \textit{Phase 2}, forming a stationary domain wall between \textit{Phase 1} and \textit{Phase 3}. In Fig.~\ref{fig8}(b), we show experimentally measured spatiotemporal maps of displacement confirming the formation of a stationary domain wall.

In Fig.~\ref{fig8}(c), we plot the transient response of several particles in the chain. We observe that the 5th and the 6th particles remain stationary in that they do not snap to neighboring stable wells, forming a stationary domain wall; however, their equilibrium state is slightly perturbed from \textit{Phase 2} due to the coexistence of other states next to them.
We also verify that stationary domain walls can form in longer chains. Moreover, the domain wall could be made of only one particle (with $u \approx 0$) if the chain consists of an odd number of particles. Refer to Supplementary Material (Section IV) for details.

To further explore the tunability of onsite potential, we vary the system parameters \(d_1\) and \(d_3\), ensuring \(d_1 = d_3\), while keeping \(d_2 = 17 \, \mathrm{mm}\) fixed, as shown in Fig.~\ref{fig9}(a). First, due to the symmetry of the onsite potential, we can analytically calculate the width of these transition waves using the \(\phi^6\) model (see Supplementary Material, Section V, for details) and plot the results in Fig.~\ref{fig9}(b). The analysis reveals that the spatial width of the colliding waves increases with increasing \(d_3\), indicating that wave width decreases as the asymmetry of the energy wells becomes larger. We then simulate the collision of \(2 \rightarrow 1\) and \(2 \rightarrow 3\) transition waves. 

Upon collision, we plot the steady-state displacements of the middle particles (4th, 5th, 6th, and 7th) in a 10-particle chain, as shown in Fig.~\ref{fig9}(c). Similarly, Fig.~\ref{fig9}(d) illustrates the steady-state displacements of the middle particles (4th, 5th, and 6th) in a 9-particle chain. We observe that, in general, more middle particles shift toward \textit{Phase~2} as \(d_3\) increases. However, in odd-particle chains, the middle particle always stabilizes at \textit{Phase~2}. Overall, the zone of influence, which defines the stationary domain wall formed after the collision of two transition waves, increases with \(d_3\). This observation aligns with the increasing wave width as \(d_3\) grows.

To quantify the width of the stationary domain wall, we define the zone of influence as the number of unit cells in their steady state within a tolerance range of \(\pm 12\) mm around \textit{Phase~2}. Figures~\ref{fig9}(e) and \ref{fig9}(f) show the size of the stationary domain wall. For even-particle chains, the size increases from zero to two particles as \(d_3\) increases, while for odd-particle chains, the size grows from one to three particles. This again highlights the correlation between the colliding wave width and the stationary domain wall width, both of which increase as \(d_3\) increases.
}

\begin{figure}[!]
    \centering
    \includegraphics[width=\linewidth]{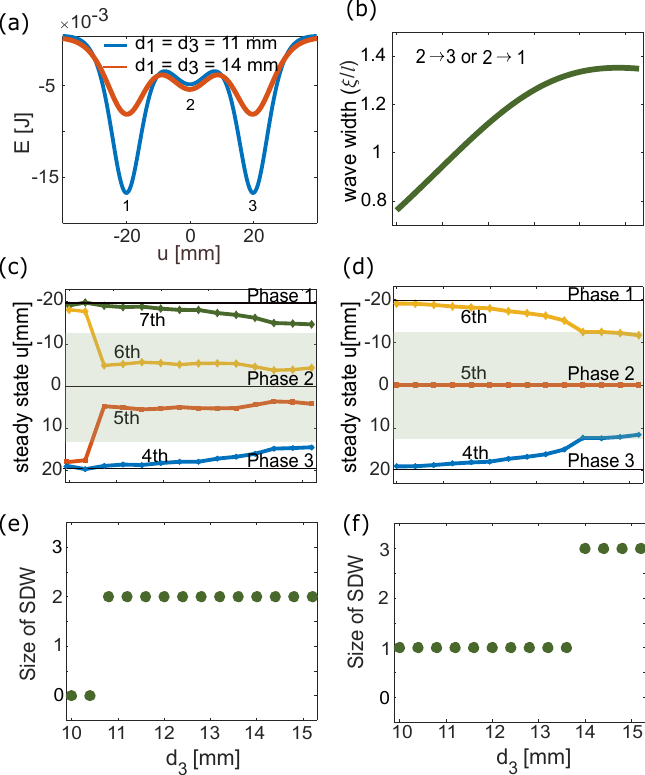}
    \caption{\justifying 
    (a) Onsite potential for varying $d_1$ and $d_3$ keeping $d_2= 17$ mm fixed. 
    (b) Wave width (normalized with respect to the unit cell length $l$) of $2\rightarrow 3$ or $2\rightarrow 1$ transition wave vs $d_3$ derived analytically. 
    (c) and (d) Steady state displacements of the middle particles in the chain for even and odd lattice. (e) and (f) Size of stationary domain wall for even and odd lattice for varying $d_3$.
    } 
    \label{fig9}
\end{figure}

Finally, similar to the ability to control the spatial location of nucleation, the location of the domain wall can be tuned by introducing a time delay in triggering the transition waves from either end (refer to Supplementary Material (Section IV) for details). 

\section{Conclusion}
In summary, we investigate a one-dimensional chain with localized external magnetic fields, formed by an assembly of permanent magnets. Specifically, our focus is on a tristable lattice, where we experimentally verify the existence of different types of transition waves. These waves sustainably propagate in the lattice due to the designed asymmetry in the potential well. We also verify experimentally a scaling law that relates the averaged power dissipated to the asymmetry in the potential well and wave velocity for all types of transition waves. We also report the reflection of transition waves from finite boundaries.

Additionally, we explore the collision of transition waves. In the case of an asymmetric potential well, when two transition waves collide as a kink and anti-kink, we observe experimentally the remote nucleation of a new phase. We report the occurrence of nucleation for a range of asymmetry in tristable potential. For larger asymmetry, nucleation is accompanied by propagation due to the large kinetic energy of nucleating particles. However, in the case of a symmetric potential well, two transition waves collide as kinks, resulting in the formation of a stationary domain wall between two different phases. We show the width of the stationary domain wall can be tuned by the shape of the tristable potential, which also dictates the width of the colliding transition waves.

These findings underscore the richness of dynamical phenomena in multistable lattices. The design holds promise for the development of reconfigurable materials under external fields, where remote actuation through transition waves can be utilized to tune the final state of the material. 

Future work will focus on extending these findings to higher-dimensional systems. Additionally, it would be intriguing to explore whether similar tunability can be achieved by modifying intersite potentials instead of onsite potentials, as investigated in this study. Preliminary analysis (see Supplementary Material, Section VI) indicates that intersite multistability leads to transitions occurring in the strains of the connecting springs rather than their displacements.


\section*{Acknowledgments}
We acknowledge helpful discussions with Prof. Hiromi Yasuda. A.R. gratefully acknowledges the support from the Institute of Eminence (IoE) IISc Postdoctoral Fellowship. 

\section*{Author Contributions}
V.D. and R.C. conceived the project. A.R. conducted the theoretical and numerical studies. S.A. fabricated the design, performed the experiments, and analyzed the data. All authors contributed to writing the manuscript. V.D. and R.C. supervised the overall project.

\def\bibsection{\section*{References}} 

\bibliography{main_ref}

\clearpage
\onecolumngrid
\setcounter{equation}{0}
\setcounter{figure}{0}
\setcounter{section}{0}
\section*{Supplementary Material}
\section{Calculation of onsite potential}
\label{Apendix_demag_energy}
A uniform magnetized sphere B, supporting a magnetization $\textbf{m}(\textbf{x})$ generates a magnetic field $\textbf{h}(\textbf{x})$ on all $R^3$. The magnetic field $\textbf{h}(\textbf{x})$ can be obtained by solving Maxwell's equation of magnetization:
\begin{equation}
\begin{split}
    \begin{aligned}
        \nabla \times \mathbf{h}(\textbf{x}) &= 0, \\
        \nabla \cdot (\mathbf{h}(\textbf{x}) + \mathbf{m} \, \chi_{_B}) &= 0,
    \end{aligned}
\end{split}
\label{appx1_eq1}
\end{equation}
where $\chi_{_B}$ is the characteristic function defined as follows:
\begin{equation}
   \chi_{_B} = \begin{cases}
  1  & \text{for } \mathbf{x} \in B \\    
  0   & \text{for } \mathbf{x} \in R^3 \texttt{\textbackslash} B\,.
\end{cases}
\label{appx1_eq2}
\end{equation}
\begin{figure*}[h]
\renewcommand{\thefigure}{S\arabic{figure}}
\centering
    \includegraphics[width=0.9\textwidth]{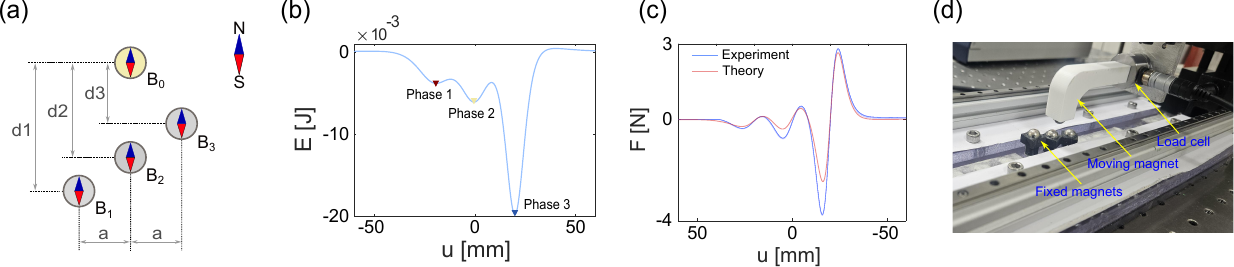}
    \caption{\justifying (a) Schematic of the unit cell, where the fixed magnets are kept at depths $d_1$, $d_2$ and $d_3$ from the the moving magnet. (b) Onsite potential of tristable unit cell. (c) Experimental and theoretical comparison of onsite force. (d) Setup showing the displacement-controlled loading experiment to measure the onsite force.}   
    \label{appx1:fig1}
\end{figure*}
For a uniformly magnetized sphere, the induced magnetic field $\textbf{h}(\textbf{x})$ is given as~\cite{james1994internal}
\begin{equation}
    \textbf{h}(\textbf{x}) = \begin{cases}
        -\dfrac{1}{3} \mathcal{I}\textbf{m}  & \text{for } \mathbf{x} \leq B \\[2em]
        -\dfrac{r^3}{3}\left(\dfrac{\mathcal{I}}{|\mathbf{x}|^3} - 3\dfrac{\mathbf{x} \otimes \mathbf{x}}{|\mathbf{x}|^5}\right)\textbf{m} & \text{for } \mathbf{x} \geq B\,.
    \end{cases}
    \label{appx1_eq3}
\end{equation}
The magnetostatic energy of a uniformly magnetized sphere can be computed as follows 
\begin{equation}
    ME= k_d \int_{R^3} |\mathbf{h}|^2\, \mathbf{dx} = -k_d \int_{R^3} \mathbf{h} .  \mathbf{m} \,\mathbf{dx} =-k_d \int_{B} \mathbf{h} .  \mathbf{m}\, \mathbf{dx}= -k_d \mathbf{m}.\int_{B} \mathbf{h}\, \mathbf{dx}  = k_d \mathbf{m}. \dfrac{\mathcal{I}}{3} \mathbf{m} |B|
    \label{appx1_eq4}
\end{equation}
where $k_d= m_s^2/2\mu_0$, $m_s$ being the magnetization of the uniformly magnetized sphere, $\mu_0 = 4 \pi \times 10^{-7} \, \text{Vs/Am}$ denotes vacuum permeability and $|B|$ denotes the volume of the sphere.

We now write down the total magnetostatic energy of four uniformly magnetized identical spheres $B_0$, $B_1$, $B_2$, and $B_3$ (see Fig. \ref{appx1:fig1}(a)) supporting magnetization $\mathbf{m}_0$, $\mathbf{m}_1$, $\mathbf{m}_2$ and $\mathbf{m}_3$ respectively, as
\begin{equation}
\begin{split}
    ME &= k_d \int_{R^3} |\mathbf{h}_0+ \mathbf{h}_1+ \mathbf{h}_2+ \mathbf{h}_3|^2 \mathbf{dx} \\
    &= \underbrace{k_d \left( \int_{R^3} {h_0}^2 \mathbf{dx} + \int_{R^3} {h_1}^2 \mathbf{dx} + \int_{R^3} {h_2}^2 \mathbf{dx}+\int_{R^3} {h_3}^2 \mathbf{dx} \right)}_{\text{self-energies}} +\\
    &\quad  \underbrace{2k_d \left( \int_{R^3} \mathbf{h}_0. \mathbf{h}_1 \, \mathbf{dx} + \int_{R^3} \mathbf{h}_0. \mathbf{h}_2\, \mathbf{dx}+\int_{R^3} \mathbf{h}_0. \mathbf{h}_3 \,\mathbf{dx}+\int_{R^3} \mathbf{h}_1. \mathbf{h}_2 \,\mathbf{dx} \right. +\left. \int_{R^3} \mathbf{h}_1. \mathbf{h}_3 \,\mathbf{dx}+\int_{R^3} \mathbf{h}_2. \mathbf{h}_3 \, \mathbf{dx}\right)}_{\text{interaction energies}}\,.
\end{split}
\label{appx1_eq5}
\end{equation}
The first four integrals represent the self energies and the last six integrals represent the interaction energies. Note that the interaction energy is of the form $\int \mathbf{h}_i. \mathbf{h}_j \,\mathbf{dx}$ and represents the interaction energy of the $i$th magnet and the $j$th magnet ($i \neq j$). This interaction energy depends on the distance between the $i$th magnet and the $j$th magnet. 

Since $B_1$, $B_2$ and $B_3$ are fixed and uniformly magnetized spheres; $\int \mathbf{h}_1. \mathbf{h}_2 \,\mathbf{dx}$, $\int \mathbf{h}_1. \mathbf{h}_3 \,\mathbf{dx}$ and $\int \mathbf{h}_2. \mathbf{h}_3 \,\mathbf{dx}$ are all constants. Also note that since $B_1$, $B_2$ and $B_3$ are uniformly magnetized, their self energies (see Eq.~\eqref{appx1_eq4}) can be written as
\begin{equation}
    \int_{R^3} {h_0}^2 \mathbf{dx} = \int_{R^3} {h_1}^2 =\mathbf{dx}\int_{R^3} {h_2}^2 \mathbf{dx}=\int_{R^3} {h_3}^2 \mathbf{dx}= k_d \mathbf{m}. \dfrac{\mathcal{I}}{3} \mathbf{m} |B|= constant\,.
    \label{appx1_eq6}
\end{equation}
Evaluation of the integral in Eq.~\eqref{appx1_eq5} yields the total magnetostatic energy of the system as follows:
\begin{equation}
    ME = \underbrace{ 2k_d \left( \int_{R^3} \mathbf{h}_0. \mathbf{h}_1 \, \mathbf{dx} + \int_{R^3} \mathbf{h}_0. \mathbf{h}_2\, \mathbf{dx}+\int_{R^3} \mathbf{h}_0. \mathbf{h}_3 \mathbf{dx} \right)}_{\text{interaction energies}} + constant\,.
\label{appx1_eq7}
\end{equation}
We thus get the total magnetostatic energy (modulo a constant) as
\begin{equation}
\begin{split}
    E &= 2k_d \left( \int_{R^3} \mathbf{h}_0. \mathbf{h}_1 \, \mathbf{dx} + \int_{R^3} \mathbf{h}_0. \mathbf{h}_2\, \mathbf{dx}+\int_{R^3} \mathbf{h}_0. \mathbf{h}_3 \mathbf{dx} \right)\,.
\end{split}
\label{appx1_eq8}
\end{equation}
For the considered unit cell with three fixed magnets spaced at a distance $a$ (i.e., their centers are located at ${x=-a,\,0,\,a}$) at depths $d_1$, $d_2$ and $d_3$ from the moving magnet (see Fig.~ \ref{appx1:fig1}(a)), we can evaluate the above integral and obtain the following:
\begin{equation}
\begin{split}
E(u,a,d_1,d_2,d_3) = & \frac{4 \pi m_s^2}{9 \mu_0} r^6 \Bigg[ \Bigg(\frac{1}{ ((u + a)^2 + d_1^2)^{\frac{3}{2}} }- \frac{3d_1^2}{((u + a)^2+ d_1^2)^{\frac{5}{2}}}\Bigg) + \left(\frac{1}{(u^2 + d_2^2)^{\frac{3}{2}}}- \frac{3d_2^2}{(u ^2+ d_2^2)^{\frac{5}{2}}}\right) \\
& + \left(\frac{1}{ ((u - a)^2 + d_3^2)^{\frac{3}{2}} }- \frac{3d_3^2}{((u - a)^2+ d_3^2)^{\frac{5}{2}}}\right)\Bigg].
\end{split}
\label{appx1_eq9}
\end{equation}
The exact energy is plotted in Fig.~\ref{appx1:fig1}(b), with numerical values: $\mu_0 = 4 \pi \times 10^{-7} \, \text{Vs/Am}$, $m_s = 0.8 \,\text{T}$, ${(d_1, d_2,d_3) = (18,15.3, 10.4)}$ mm, $r=5$ mm and $a= 20$ mm. Therefore, we achieve a tristable onsite energy potential, and define the three stable states as \textit{Phase 1}, \textit{Phase 2}, and \textit{Phase 3}. The tunable depths  $d_1$, $d_2$ and $d_3$ help us to tailor the energy landscape. Fig.~\ref{appx1:fig1}(c) shows the comparison of the experimental and theoretical onsite force-displacement relation. The experimental measure of this nonmonotonous force-displacement relation is obtained by performing a displacement-controlled loading experiment and measuring the force values using a load cell, as shown in Fig.~\ref{appx1:fig1}(d). The slight deviation of the experimental curve from the theoretical one can be attributed to variations in the performance of the N-52 magnets. These variations occur because the magnets are manufactured in different blocks and batches, which themselves can vary. Specifically, the variations in different blocks and batches may be due to differences in the size of NdFeB powders or the thermal history each magnet experiences while being sintered in a furnace. Additionally, variations can also arise from slight differences in the microstructure of the starting NdFeB ingot, from which the powders are generated\cite{cui2022manufacturing}.
\begin{figure*}[!]
\renewcommand{\thefigure}{S\arabic{figure}}
\centering
    \includegraphics[width=0.76\textwidth]{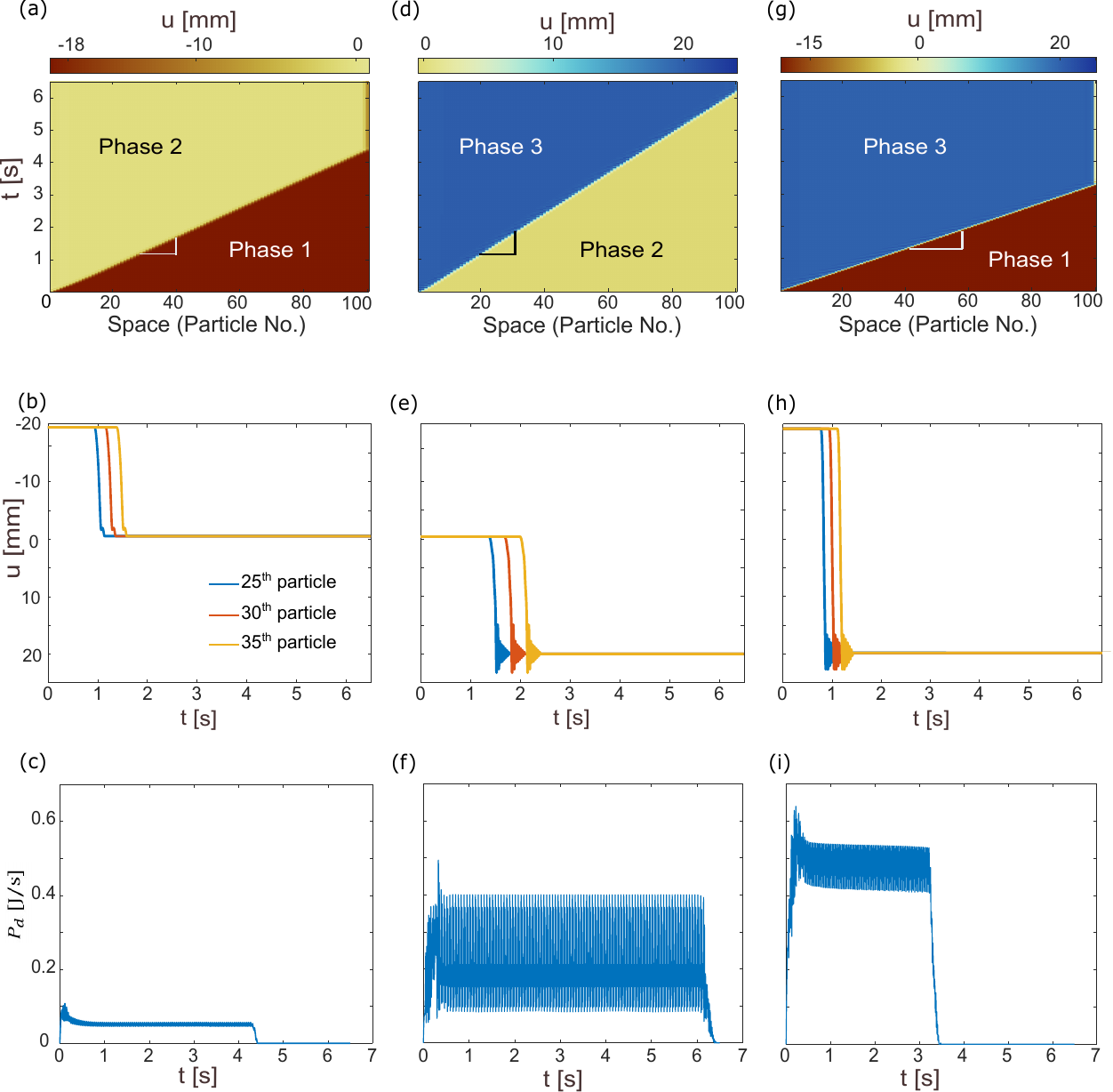}   
     \caption{\justifying Numerical study for a chain of 100 particles. (a) (d) (g) Spatiotemporal plot of displacement (b) (e) (h) displacement-time series, and (c) (f) (i) Dissipated power, for (a) (b) (c) $1 \rightarrow 2$ transition, (d) (e) (f) $2 \rightarrow 3$ transition, (g) (h) (i) $1 \rightarrow 3$ transition.}  
    \label{appx2:fig3}
\end{figure*}
\section{Transition waves in longer chains with damping}
Considering the system parameters used in experiments, we conduct numerical simulations on longer chains consisting of 100 particles. With similar initial conditions as described in the main text, we observe that all $1 \rightarrow 2$, $2 \rightarrow 3$, and $1 \rightarrow 3$ transition waves are sustainable even in longer chains, as shown in Fig. \ref{appx2:fig3}.

We also calculate the dissipated power $(P_d)$ for all three transitions. We observe that the dissipated power oscillates in time with a period $T=l/\nu$. We compute the time-averaged $P_d$ and utilize it to verify the scaling law (Eq. 11) numerically and experimentally (as seen in Fig. 3(e) in the manuscript).




\section{Nucleation in longer chains with damping}
\begin{figure}[!]
\renewcommand{\thefigure}{S\arabic{figure}}
\centering
    \includegraphics[width=0.72\textwidth]{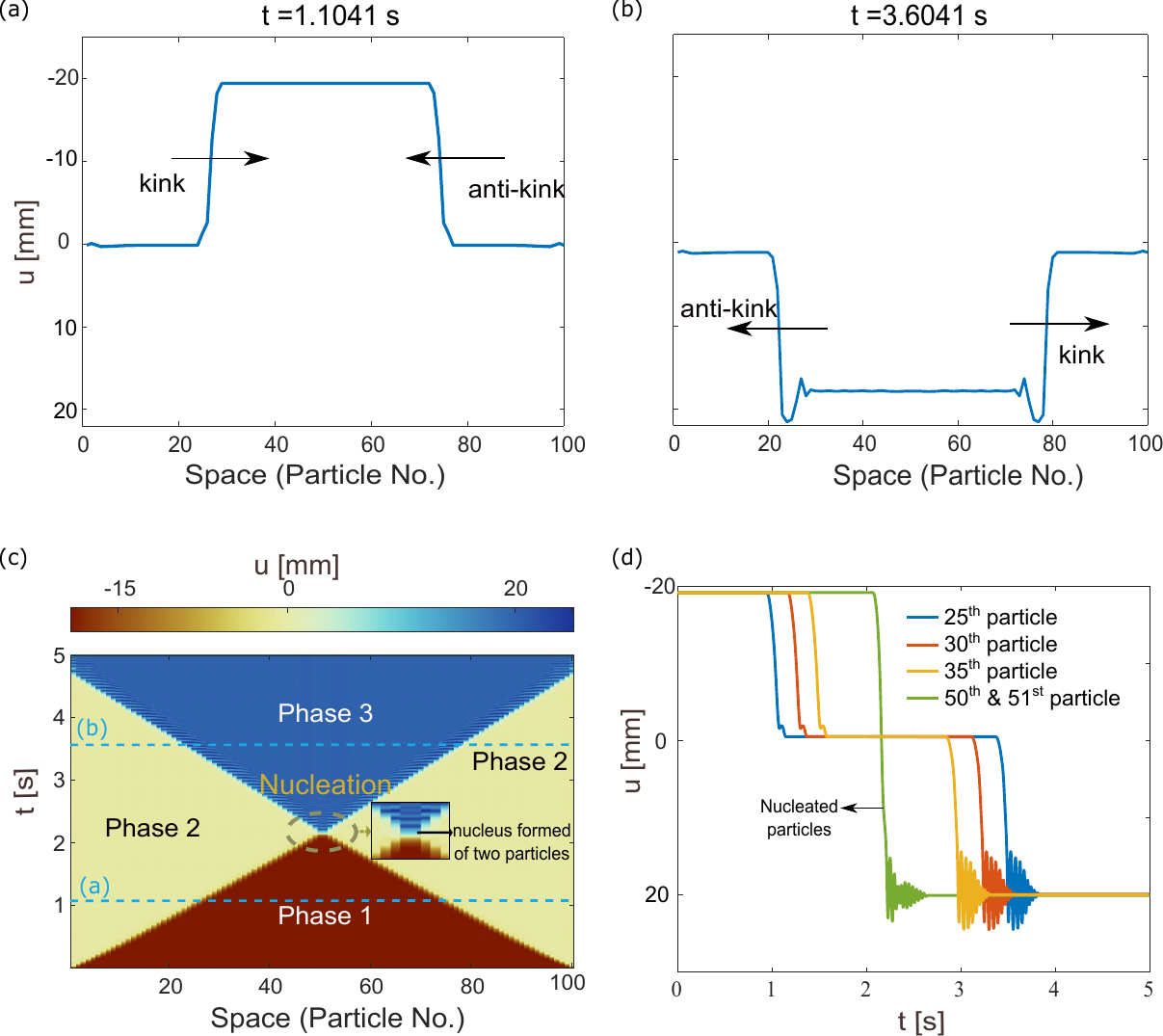}
    \caption{\justifying Collision of a kink and and an anti-kink resulting in the formation of a new phase in longer chains. (a)-(b) spatial profiles at time 1.1041 s and 3.6041 s respectively. (c) Spatiotemporal plot of displacement for the formation of a \textit{Phase 3} upon collision of two transition fronts $1 \rightarrow 2$, (d) displacement-time series.} 
    \label{appx2:fig4}
\end{figure}

We investigate longer chains comprising 100 unit cells for nucleation study. Initially, all unit cells are in \textit{Phase 1}. $1 \rightarrow 2$ transition waves are triggered from opposite ends. Despite damping, these waves persist in longer chains, as shown in Fig.~\ref{appx2:fig4}(a).
The collision of the kink and anti-kink occurs at the center, leading to their annihilation and the emergence of a new phase (\textit{Phase 3}). This new phase initiates the formation of a new pair of kink and anti-kink, propagating in opposite directions as $2 \rightarrow 3$ transition waves shown in Fig.~\ref{appx2:fig4}(b).
Thus, the current configuration, with asymmetric potentials and the collision dynamics of kink and anti-kink, presents a robust method for remotely nucleating a new phase even in longer chains and damped systems.

\begin{figure}[!]
\renewcommand{\thefigure}{S\arabic{figure}}
\centering
    \includegraphics[width=0.72\textwidth]{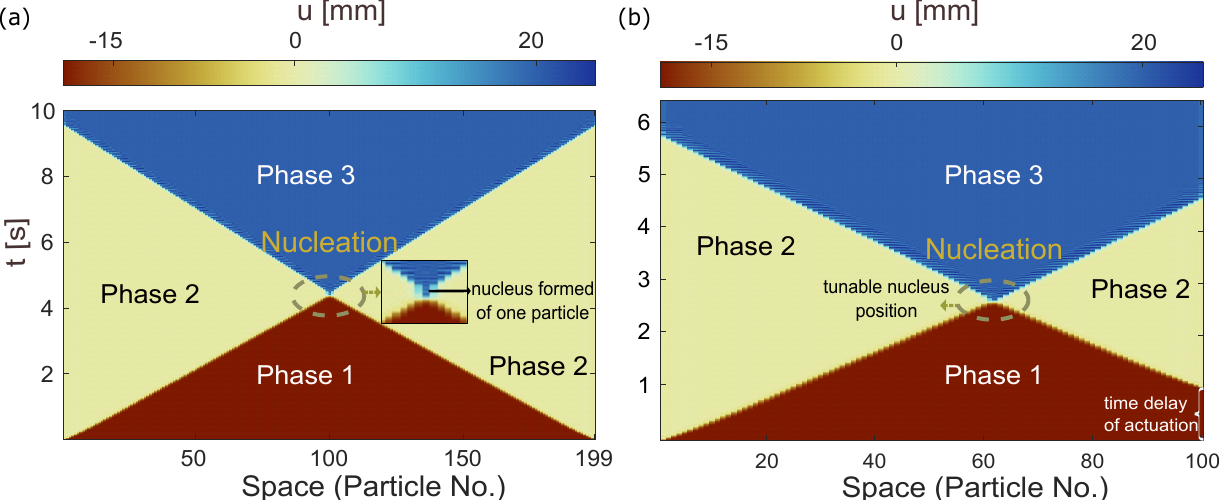}
    \caption{\justifying  Nucleation observed when lattice is made up of odd number of unit cells (199 particles). (a) Spatiotemporal plot of displacement for 199 particles. Tunable spatial location of nucleation using time delay of actuation, (b) spatiotemporal plot of displacement  when the right end of the chain is actuated at delay of 1 s.} 
    \label{appx2:fig5}
\end{figure}

Note that the nucleus of the new phase shown in Figs.~\ref{appx2:fig4}(c),(d) consists of the two middlemost particles in chains with an even number of unit cells. However, for an odd number of particles ($n=199$) in the chain, the collision of a kink and anti-kink from either side of the chain results in only the central (singular) element acquiring \textit{Phase 3}. The nucleus formed by this single element (the 100th particle) effectively initiates the propagation of the new phase in both directions of the chain, as shown in Fig.~\ref{appx2:fig5}(a).

Moreover, we can tailor the spatial location of nucleation using a time delay of actuation from either end, as demonstrated in Fig.~\ref{appx2:fig5}(b). We consider a chain consisting of 100 unit cells. We apply similar initial conditions at both ends (\textit{Phase 1} to \textit{Phase 3}), but the right end of the chain is actuated with a delay of 1 s. Consequently, a nucleus is formed upon the collision of two waves, but not precisely at the center of the chain. Instead, the nucleus is biased towards the right end of the chain. Nonetheless, this nucleus can effectively trigger $2 \rightarrow 3$ transition waves in both directions of the chain. Hence, the time-delayed actuation from either end provides a convenient method to tailor the spatial location of remote nucleation within the lattice.

\section{Stationary domain walls in longer chains with damping}
\begin{figure*}[!]
\renewcommand{\thefigure}{S\arabic{figure}}
\centering
    \includegraphics[width=0.73\textwidth]{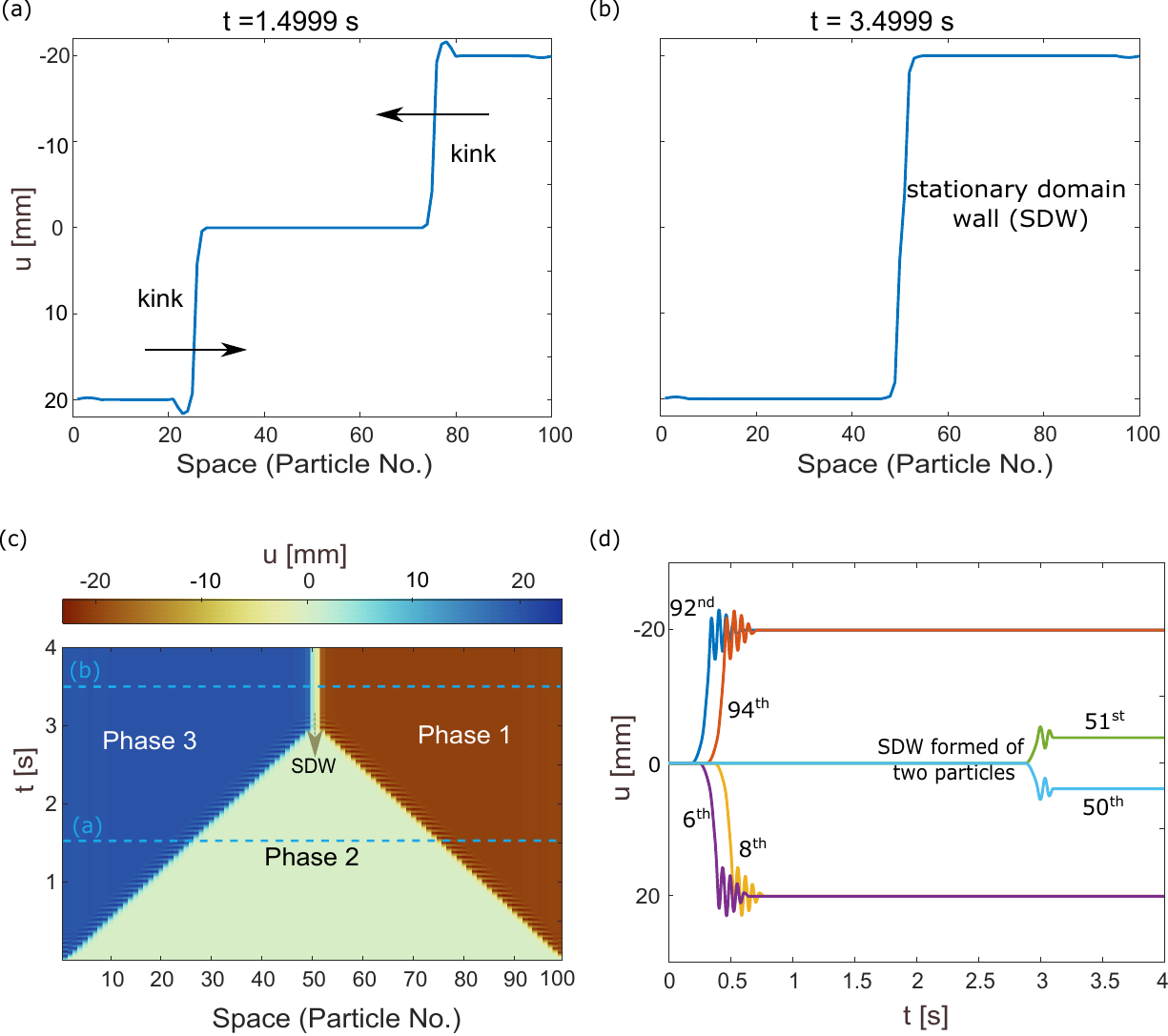}
    \caption{\justifying Collision of two kinks resulting in a stationary domain wall in longer chains. (a)-(b) spatial profiles at time 1.4999 s and 3.4999 s respectively. (c) spatiotemporal plot of dsplacement and (d) displacement time series, for the case when the chain is constituted of even (100) number  of particles. The stationary domain wall is formed of two unit cells. } 
    \label{appx2:fig6a}
\end{figure*}

\begin{figure}[!]
\renewcommand{\thefigure}{S\arabic{figure}}
\centering
    \includegraphics[width=0.72\textwidth]{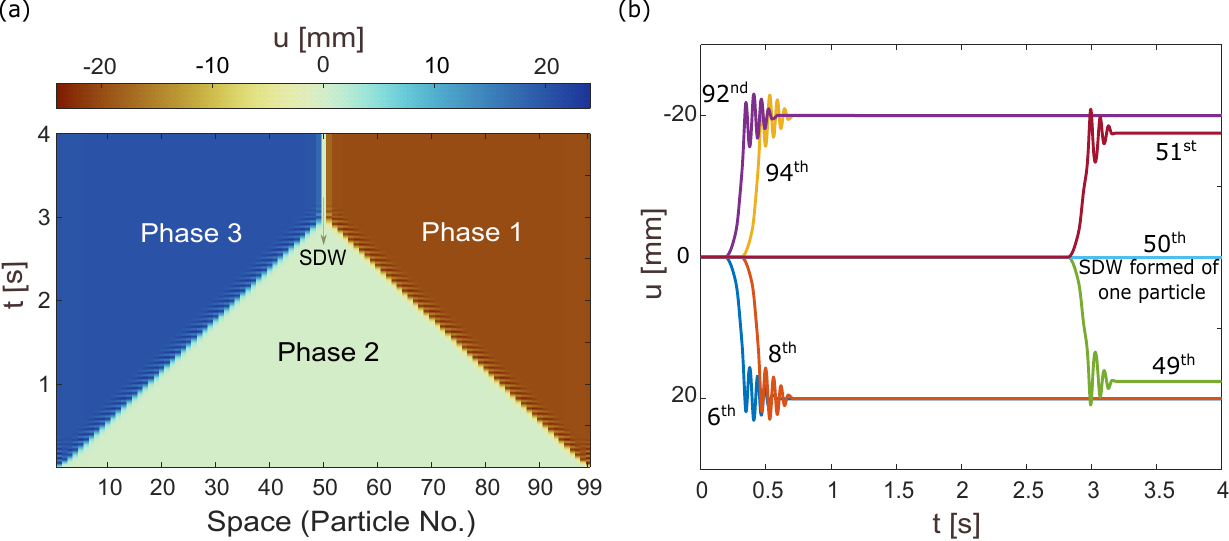}
    \caption{\justifying Formation of domain walls in case of odd number of particles (a) spatiotemporal plot of displacement and (b) displacement time series for the case when the chain is constituted of odd (99) number of particles. The stationary domain wall is formed of one unit cell.} 
    \label{appx2:fig6b}
\end{figure}

We perform numerical simulations on longer chains of length 100 (and 99) with opposite initial conditions, i.e. we trigger $2 \rightarrow 3$ and $2 \rightarrow 1$ transition waves from opposite ends. Upon collision of two kinks, we clearly observe the formation of domain wall in both cases, as shown in Figs. \ref{appx2:fig6a} and \ref{appx2:fig6b}. For odd chain lengths, a single particle forms the stationary domain wall, while in even chain lengths, the stationary domain wall is shared by two particles. Note that the steady state of the stationary domain wall is different in odd and even number of particles. When we have a chain made of odd number of unit cells, the stationary domain wall remains exactly at \textit{Phase 2} $(u \approx 0)$, as shown in Fig.~\ref{appx2:fig6b}(b). However, when the number of elements in the chain is even, the domain wall is shared by the middlemost two elements, having their equilibrium state slightly perturbed from \textit{Phase 2}, as shown in Fig.~\ref{appx2:fig6a}(d).  

\begin{figure}[!]
\renewcommand{\thefigure}{S\arabic{figure}}
\centering
    \includegraphics[width=0.72\textwidth]{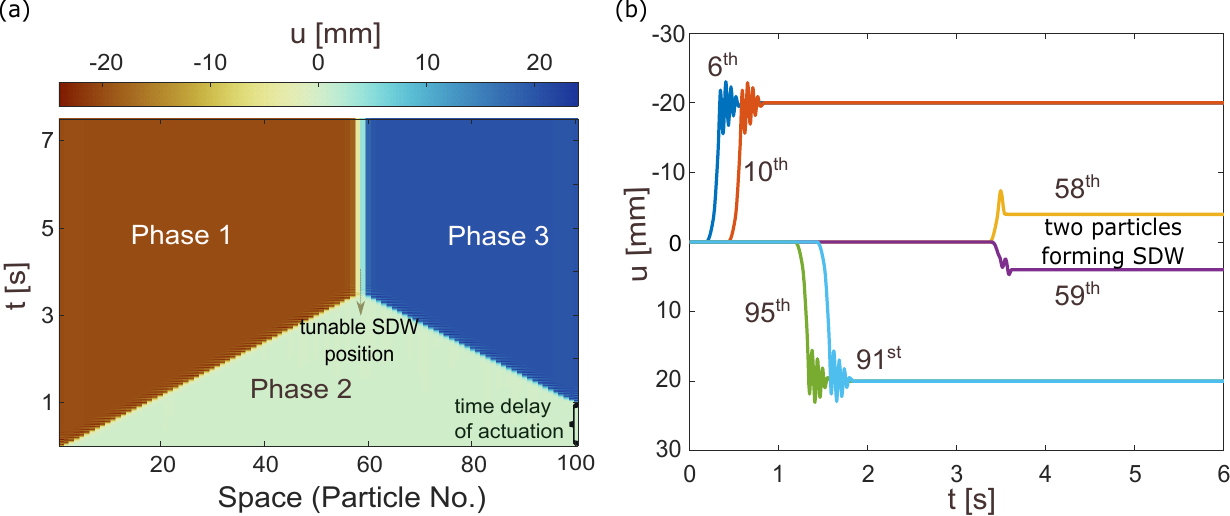}
    \caption{\justifying Spatially tunable domain wall. The right end of the chain is actuated at a delay of 1 s. (a) spatiotemporal plot of displacement and (b) displacement time series of 100 particles. The SDW is formed of 58th and 59th particle.}   
    \label{appx2:fig7}
\end{figure}

Furthermore, by adjusting the timing of actuation from each end, as illustrated in Fig.~\ref{appx2:fig7}, we can customize the spatial position of the stationary domain wall. We examine a chain comprising 100 unit cells. While triggering $2 \rightarrow 1$ and $2 \rightarrow 3$ transition waves from opposite ends, we introduce a 1-second delay in actuation at the right end of the chain. Consequently, upon the collision of these waves, a stationary domain wall forms, not exactly at the chain's center. Instead, it biases towards the right end. Thus, employing time-delayed actuation from either end offers a convenient means to adjust the spatial location of stationary domain wall within the lattice.

\section{Analytical solution for transition wave in case of symmetric tristable potential}
\begin{figure}[t!]
    \centering
    \includegraphics[width=0.8\linewidth]{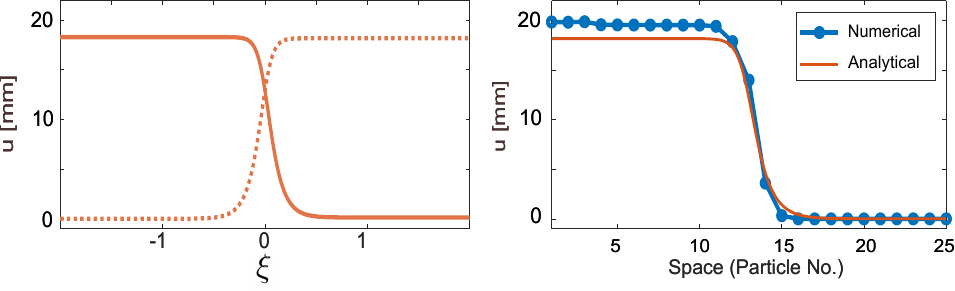}
    \caption{\justifying (a) Analytical solution of kink (dotted) and anti-kink (solid) type waveform in the traveling coordinate $\xi$, assuming $\phi^6$ model approximation. (b) Waveform at $t=0.78$ s obtained from numerical simulation (blue curve) and analytical solution.
     } 
    \label{R9}
\end{figure}
For conservative system the equations of motion are
\begin{eqnarray}
\begin{array}{ll}
m{u_{n,tt}}-k(u_{n-1}-2u_n+u_{n+1})+E'(u_n)=0.
\end{array}
\label{A_eq10}
\end{eqnarray}
We introduce traveling coordinate $\xi= nl-\nu t$ and assume $u_n(t)\equiv u(nl-\nu t) \equiv u(\xi)$ and substitute it to Eq.~\eqref{A_eq10} to yield
\begin{eqnarray}
\begin{array}{ll}
m\nu^2 u_{,\xi \xi}-k\big(u(\xi-l)-2u(\xi)+u(\xi+l)\big)+E'(u)=0.
\end{array}
\label{A_eq11}
\end{eqnarray}
Using Taylor's series expansions the displacements $u(\xi\pm l)$ can be expressed as
\begin{equation}
   u(\xi\pm l) = u(\xi) \pm l\dfrac{\partial u}{\partial \xi}+ \dfrac{l^2}{2!}\dfrac{\partial^2 u}{\partial \xi^2} \pm \dfrac{l^3}{3!} \dfrac{\partial^3 u}{\partial \xi^3} + h.o.t.
    \label{A_eq12}
\end{equation}
Retaining only the terms up to quadratic order of Eq. \eqref{A_eq12} and neglecting the higher order terms for the continuum limit (small $l$), Eq. \eqref{A_eq11} reduces to
\begin{equation}
 u_{,\xi \xi}= \dfrac{E'(u)}{m(c_0^2l^2-\nu^2)},
 \label{A_eq13}
\end{equation}
where $c_0^2= k/m$. For tristable wells symmetric about the $y$-axis, we rewrite the right hand side of Eq.~\eqref{A_eq13} such that
\begin{equation}
u_{,\xi \xi}= \eta_1 u - \eta_3 u^3 + \eta_5 u^5.
 \label{A_eq14}
\end{equation}
Equation \eqref{A_eq14} has the form of a nonlinear Klein Gordan equation with $\phi^6$ model potential. The model admits to the following analytical solution \cite{wazwaz2006compactons}: 
\begin{equation}
u(\xi) = \sqrt{\frac{2\eta_1}{\eta_3} 
{ \left ( 1\pm \tanh(\sqrt{\eta_1}\xi) \right )} },
    \label{A_eq15}
\end{equation}
with the constraint
\begin{equation}
 \eta_5 = \dfrac{3 \eta_3^2}{16 \eta_1}.
    \label{A_eq16}
\end{equation}

Figure~\ref{R9}(a) illustrates the waveform of the analytical solution as obtained from Eq.~\eqref{A_eq15}. The tristable onsite wells are considered with $(d_1, d_2, d_3)= (14, 15, 14)$ mm. The $2 \rightarrow 3$ transition wave velocity is derived from the corresponding numerical simulation, $\nu \approx 1.5\, m/s$.  Figure~\ref{R9}(b) shows the waveform of the numerical simulation with that of the analytical solution. The analytical solution exhibits a qualitatively similar profile, particularly in the width of the transition wave, which is defined as $\dfrac{\ln (3)}{\sqrt\eta_1}$ based on the half maximum value.



\section{Intersite Potentials}
\begin{figure}[b!]
\renewcommand{\thefigure}{S\arabic{figure}}
\centering
    \includegraphics[width=0.9\textwidth]{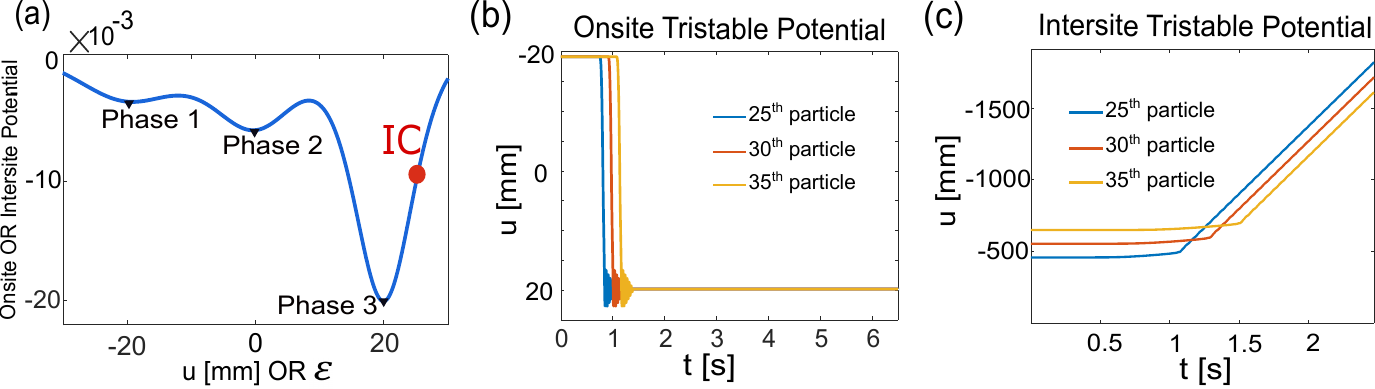}
    \caption{\justifying (a) Potential function representing  onsite/intersite tristability. Initial condition (IC) is marked. Displacement-time profile of several particles in the chain which possess (b) onsite tristability; (c) intersite tristability.} 
    \label{R7}
\end{figure}

\begin{figure}[b!]
\renewcommand{\thefigure}{S\arabic{figure}}
\centering
    \includegraphics[width=0.65\textwidth]{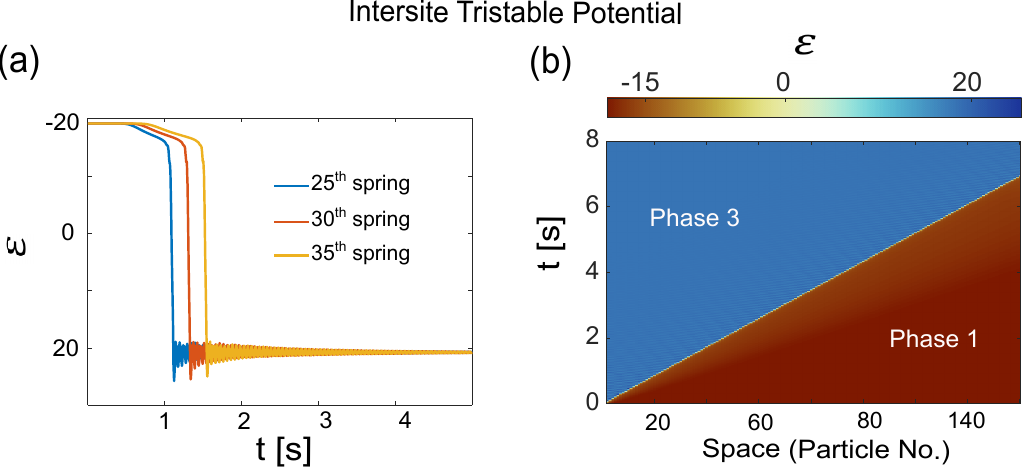}
    \caption{\justifying (a) Stain-time profile of several interconnecting springs in the chain. We see the strains of the springs undergoing a transition from $1\rightarrow 3$. (b) Spatiotemporal map of strain, showing transition from \textit{Phase 1}-strain to \textit{Phase 3}-strain.}
    \label{R8}
\end{figure}
To explore the effect of intersite potentials, let us consider a system with masses connecting to its neighbors via springs. But in this case, we do not consider onsite fixed magnets that contribute to onsite tristability. Instead, the interconnected springs are non-linear and exhibit a similar tristable potential as defined in Eq.~(1) in the manuscript, i.e., the strain energy of the spring is defined as
\begin{equation}
\begin{split}
E(\varepsilon) = & \frac{4 \pi m_s^2}{9 \mu_0} r^6 \Bigg[ \Bigg(\frac{1}{ ((\varepsilon + a)^2 + d_1^2)^{\frac{3}{2}} }- \frac{3d_1^2}{((\varepsilon + a)^2+ d_1^2)^{\frac{5}{2}}}\Bigg)  + \left(\frac{1}{(\varepsilon^2 + d_2^2)^{\frac{3}{2}}}- \frac{3d_2^2}{(\varepsilon ^2+ d_2^2)^{\frac{5}{2}}}\right) \\
& + \left(\frac{1}{ ((\varepsilon - a)^2 + d_3^2)^{\frac{3}{2}} }- \frac{3d_3^2}{((\varepsilon - a)^2+ d_3^2)^{\frac{5}{2}}}\right)\Bigg],
\end{split}
\label{A_eq17}
\end{equation}
where $\varepsilon$ is strain of the spring. 
For instance, if the $n$th spring links two masses having displacements $u_{n-1}$ and $u_{n}$, then the strain of the $n$th spring is expressed as
\begin{equation} 
\varepsilon_{n} = u_{n} - u_{n-1}. 
\label{A_eq18}
\end{equation}
Equation (\ref{A_eq17}) characterizes intersite tristability, i.e., each interconnecting spring can exist in any one of three stable states of strain: \textit{Phase 1}, \textit{Phase 2}, or \textit{Phase 3} as shown in Fig. \ref{R7}(a). For instance, if the $n$th spring is in \textit{Phase 1}, then $\varepsilon_{n}$ corresponds to the abscissa of \textit{Phase 1}.

We write the equations of motion for the $n$th unit cell having non-linear neighbouring interactions and intersite damping as
\begin{equation}
\begin{split}
& m u_{n,tt}  + E'(u_{n-1} - u_n) - E'(u_n-u_{n+1})  - c_1(u_{(n-1),t} - 2u_{n,t} + u_{(n+1),t})  = 0,
\end{split}
\label{A_eq19}
\end{equation}
where $m$ represents the mass, $E'(\varepsilon)$ denotes the intersite force due to intersite potential $E(\varepsilon)$, $c_1$ is intersite damping parameter. Variables following a comma in indices denote partial derivatives.

For our analysis, we examine a chain comprising 160 unit cells with 161 springs, fixed at both ends. In this preliminary study, we focus solely on the investigation of the $1\rightarrow 3$ transition. We first simulate the chain with only onsite potentials (as in the main manuscript) and initial conditions as discussed in Section IV-A ($1\rightarrow 3$). The displacement profile (Fig.~\ref{R7}(b)) in this case shows transitions in displacements ($1\rightarrow 3$ transition). Next, to start the simulations with intersite potentials,  we assume that all the springs are strained such that they are in the first stable state: \textit{Phase 1}. To implement this, we ensure that for all $n$,
$\varepsilon_n$ equals the abscissa of \textit{Phase 1}. Next, to introduce a perturbation, we enforce a different strain in the first spring by displacing the left boundary to $-25.5$ mm. 
The initial strain (or initial condition) thus induced in the first spring can be seen in Fig.~\ref{R7}(a). This causes a force imbalance between the first two springs. The simulations forefront that the displacement profile of the moving masses show a distinct absence of transition/kink characteristics as deciphered from Fig.~\ref{R7}(c). However, the initial condition initiates tension in the first spring, leading it into \textit{Phase 3}. Moreover, we observe a propagating wave, transitioning each spring from \textit{Phase 1}- strain to \textit{Phase 3}- strain as shown in Fig. \ref{R8}(a-b). In Fig.~\ref{R8}(a), we plot the transient response of strains of several springs in the chain. Fig.~\ref{R8}(b) denotes the spatiotemporal plot of strain for all springs in the chain. It is evident that we see $1\rightarrow 3$ strain transition in springs possessing intersite tristability. Throughout the simulation, we maintain the strain in the first spring as highlighted in Fig.~\ref{R7}(a). 

To validate our strain transition analysis we perform the following: the equations of motion for the $(n-1)$th unit cell is written as
\begin{equation}
\begin{split}
& m u_{{n-1},tt}  + E'(u_{n-2} - u_{n-1}) - E'(u_{n-1}-u_{n})  - c_1(u_{(n-2),t} - 2u_{{n-1},t} + u_{n,t})  = 0,
\end{split}
\label{A_eq20}
\end{equation}
Subtracting Eq. (\ref{A_eq20}) from Eq. (\ref{A_eq19}) we get
\begin{equation}
\begin{split}
& m \varepsilon_{{n},tt}  + 2E'(-\varepsilon_{n}) - E'(-\varepsilon_{n+1})- E'(-\varepsilon_{n-1}) - c_1(\varepsilon_{(n-1),t} - 2\varepsilon_{{n},t} + \varepsilon_{(n+1),t})  = 0,
\end{split}
\label{A_eq21}
\end{equation}

Equation (\ref{A_eq21}) has the same structure as Eq. (1) in the manuscript, but it pertains in strain rather than displacement. Therefore, we conclude that the primary distinction between onsite and intersite potential lies in the observation that in onsite multistability, transitions occur in displacements (Fig.~\ref{R7}(b)); whereas in intersite multistability, transitions occur in strains of the interconnecting springs (Fig.~\ref{R8}(a-b)).
\end{document}